\documentclass{aa}
\usepackage{txfonts}
\usepackage{natbib}
\usepackage{float}
\usepackage{textcomp, gensymb}
\usepackage{enumitem}
\usepackage{lscape}
\usepackage{array}
\usepackage{graphicx}
\usepackage[normalem]{ulem}
\usepackage[dvipsnames]{xcolor}
\renewcommand{\arraystretch}{1.5}

\begin{document}
\bibpunct{(}{)}{;}{a}{}{,}

\title{Enlargement of depressions on comet 81P/Wild 2: Constraint based on 30-year cometary activity in the inner Solar System}

\author{Bumhoo Lim\inst{\ref{inst1}, \ref{inst2}} \and Masateru Ishiguro\inst{\ref{inst1}, \ref{inst2}}}
\institute{Department of Physics and Astronomy, Seoul National University, 1 Gwanak-ro, Gwanak-gu, Seoul 08826, Republic of Korea\label{inst1}
\and
SNU Astronomy Research Center, Department of Physics and Astronomy, Seoul National University, 1 Gwanak-ro, Gwanak-gu,
Seoul 08826, Republic of Korea\\\email{bumhoo7@snu.ac.kr, ishiguro@snu.ac.kr}\label{inst2}}

\date{Received ; accepted}

\abstract{The Stardust flyby mission to Jupiter-family comet (JFC) 81P/Wild 2 (hereafter, 81P) captured its dense quasicircular depressions. Nevertheless, the formation mechanism remains a subject of ongoing debate.}
{We aimed to study how cometary activity contributed to the formation and enlargement of these depressions by analyzing Stardust flyby images and ground-based observation data.}
{We calculated the time-dependent water production rate of 81P inside the snow line ($<$3 au) and compared it with the observational data. In addition, we estimated the fallback debris mass using an observation-based model, where a dust ejection from 81P was considered to reproduce ground-based observations of the dust tail. We compared the total excavated volume of water and dust with the total depression volume derived using the 81P shape model.}
{We find that the total excavated volume after 81P was injected into the inner Solar System accounts for up to only 30 \% of the depression volume. This insufficiency suggests that a large portion ($>$70 \%) of the depressions had already existed before the comet was injected into the current orbit. In addition, we estimate the dust-to-ice mass ratio for 81P to be 2--14.}
{We suggest that most depressions were formed in the source region of the Kuiper-belt objects.}


\titlerunning{Enlargement of depressions on comet 81P/Wild 2}
\maketitle

\section{Introduction}\label{sec_introduction}

Comets are among the most primitive remnants and are considered the key objects for studying the early stages of our Solar System. However, even the comets currently observed are not genuinely pristine but rather have been continuously altered from their origins due to collisional processes \citep[e.g.,][]{Stern1995AJ....110..856S}, temporal or periodic outbursts \citep[e.g.,][]{Hartmann1993Icar..104..226H}, or loss of volatiles \citep[e.g.,][]{Prialnik2004come.book..359P}. Since these evolutionary processes leave traces on the cometary surfaces, it is important to explore the geological features of comets.

As of December 2024, six periodic comets have been explored by space projects (1P/Halley by Giotto \citealt{Keller1987A&A...187..807K}; 9P/Tempel 1 by Deep Impact \citealt{AHearn2005Sci...310..258A} and Stardust-NexT \citealt{Thomas2013Icar..222..453T}; 19P/Borrelly by Deep Space 1 \citealt{Soderblom2002Sci...296.1087S}; 67P/Churyumov-Gerasimenko by Rosetta \citealt{Thomas2015Sci...347a0440T}; 81P/Wild 2 by Stardust \citealt{Brownlee2004Sci...304.1764B}; and 103P/Hartley 2 by EPOXI \citealt{Thomas2013Icar..222..550T}; hereafter, these comets are referred to as 1P, 9P, 19P, 67P, 81P, and 103P). From images taken by onboard cameras, the comet nuclei showed remarkable diversity in shape. This observational evidence may imply that their origins and modification processes were complex, multiphase, and occurred in various directions \citep{Cheng2013Icar..222..808C}. The high-resolution images also revealed diverse geological substructures on their surfaces, such as quasicircular depressions \citep{Brownlee2004Sci...304.1764B, Belton2013Icar..222..477B, Vincent2015Natur.523...63V}, terraces \citep{Basilevsky2006P&SS...54..808B}, mesas \citep{BruckSyal2013Icar..222..610B}, and pinnacles \citep{Basilevsky2006P&SS...54..808B, Basilevsky2017P&SS..140...80B}. These various substructures are thought to have distinct origins linked to the physical properties, heterogeneity, or seasonal evolution of their parent bodies \citep[for a comprehensive review, refer to][]{ElMaarry2019SSRv..215...36E}.

Quasicircular depressions are prevalent substructures found in all resolved cometary surfaces. Distinct from ordinary impact craters with a bowl-like morphology \citep{Melosh1980AREPS...8...65M}, cometary depressions commonly show flat bottoms and steep walls without rims, implying that they were heavily modified from the primitive state \citep{Basilevsky2006P&SS...54..808B, Cheng2013Icar..222..808C}. As a result, there is little consensus regarding their initial formation, although several hypotheses have been proposed. For example, \citet{Brownlee2004Sci...304.1764B} first identified depressions on the 81P surface and suggested that they might have resulted from direct impact cratering. However, subsequent investigations have consistently indicated that the commonly observed morphology of the cometary depressions, featuring flat floors and steep walls, cannot be attributed solely to the impact process \citep{Basilevsky2006P&SS...54..808B, Holsapple2007Icar..187..345H, Vincent2015P&SS..107...53V}. \citet{Belton2013Icar..222..477B} proposed an alternative idea: "mini-outburst" events that were observed during the Stardust-NexT encounter of 9P could account for the formation of approximately 96 \% of the observed pits. After Rosetta discovered similar features on 67P, \citet{Vincent2015Natur.523...63V} stated that sinkhole collapse could have caused the formation of some of the large and deep circular depressions. This necessitated pre-existing subsurface cavities of similar depths ($d\sim200\ \rm{m}$), and a numerical approach was also used to investigate this scenario. Considering the thermal history of 67P, \citet{Mousis2015ApJ...814L...5M} conducted a numerical simulation of the sublimation of crystalline ices. They conclude that such large cavities could only form if a dust mantle was not present on the surface, leading to the assertion that current illumination conditions alone cannot fully account for the large depressions observed on the comets. \citet{Mousis2015ApJ...814L...5M} argued alternatively that other thermal processes, such as clathrate destabilization or crystallization of amorphous ices, can explain such cavities of the observed depressions. In this paper, we use the geological term "depressions" to describe the excavated quasicircular features seen on comets, which have also been referred to as "craters" \citep[e.g.,][]{Vincent2015P&SS..107...53V} or "pits" \citep[e.g.,][]{Benseguane2022A&A...668A.132B} in some of the literature.

Although the question of the formation process is still open, recent investigations by Rosetta have demonstrated a shared evolutionary trajectory of depressions on 67P under the influence of cometary activity. The major result of these studies has been that sublimation-driven erosion continually obliterates depressions by preferentially expanding their lateral dimensions through cliff retreatment rather than by deepening them \citep{Marschall2016A&A...589A..90M, Attree2018A&A...610A..76A, Benseguane2022A&A...668A.132B, GuilbertLepoutre2023PSJ.....4..220G}. Episodic cliff collapses from sudden outbursts or jets are also attributed to this expansion \citep{Vincent2016MNRAS.462S.184V, Vincent2016A&A...587A..14V, Attree2018A&A...610A..76A}.

From a broader perspective, this tendency implies a common surface evolutionary sequence that predominantly smooth comets throughout sublimation activity. For example, \citet{Vincent2017MNRAS.469S.329V} compared the general trend of surface topography for four Jupiter-family comets (JFCs; 9P, 67P, 81P, and 103P) and find that large cliff is predominantly found on the younger surface (67P and 81P). Similar evolutionary sequences have been proposed by ground-based observation, such as the correlation of surface ages with the phase function and albedo \citep{Kokotanekova2017MNRAS.471.2974K, Kokotanekova2018MNRAS.479.4665K} or the frequency of mini-outburst \citep{Kelley2021PSJ.....2..131K}.

Recent investigations have revealed the clear evolutionary path of 67P depressions under the influence of sublimation activity: the depressions are enlarged and eventually smoothed. However, little is known about the evolution of surface depressions under the influence of sublimation activity for comets other than 67P. If the hypothesis of "depression enlargement" holds for other JFCs beyond 67P, it is possible to broaden our investigation and revisit other JFCs that have already been explored. Among them, 81P particularly draws our attention because nearly the entire explored side of the nucleus was covered with quasicircular depressions \citep{Brownlee2004Sci...304.1764B}. This abnormally high depression density is not fully explained by a single formation process.

Specifically, numerical approaches by \citet{GuilbertLepoutre2023PSJ.....4..220G} determined the erosion rate of depressions on three JFCs (9P, 81P, and 103P) and suggest that water sublimation-driven erosion plays a role in depression enlargement, contributing approximately a few percent of the overall excavated features. Nevertheless, there is still a lack of observational constraints to support this numerical approach. \citet{GuilbertLepoutre2023PSJ.....4..220G} estimated the erosion rates of JFCs based on the surface properties constrained by the Rosetta measurement for 67P \citep{Benseguane2022A&A...668A.132B}. However, the surface active fraction of 81P \citep[$\sim$10 \%;][]{Sekanina2003JGRE..108.8112S} is expected to be an order of magnitude greater than that of 67P \citep[$\sim$1 \%;][]{Marschall2016A&A...589A..90M}, which may imply ten times faster surface erosion. Indeed, the typical depression sizes on 81P were nearly an order of magnitude larger than those on other JFCs, exceeding the empirical saturation limit of impact crater densities \citep{Basilevsky2006P&SS...54..808B}. This paper attempts to determine whether this diversity in depression is related to the past activity of 81P. Notably, the 81P depressions had similar traits to those of other JFCs in terms of morphology and power slope of the size distribution, implying common formation mechanisms among these depressions \citep{Thomas2013Icar..222..453T, Ip2016A&A...591A.132I}. 

81P is a typical JFC ($a=3.45\ \rm{au}$, $e=0.537$, $i=3.24\degree$, and $P=6.4\ \rm{yr}$) and was first discovered in the 1978 perihelion approach \citep{Sekanina1985AJ.....90.2335S}. It is thought to have been injected into the current JFC orbit by a very close encounter with Jupiter ($\sim$0.0061 au) in 1974. This event dramatically altered the perihelion distance from $q = 5\ \rm{au}$ to $q = 1.69\ \rm{au}$, dragging the body inside the water activity region. Since then, 81P has remained in its current orbit for five periods of perihelion (1978, 1984, 1990, 1997, and 2003) until the Stardust encounter in 2004. Although there were slight changes in the perihelion distance between 1984 and 1990 ($q = 1.7\ \rm{au}\rightarrow \it q \rm =1.6\ \rm{au}$) due to a moderate Jupiter encounter in 1988, the orbit of 81P during this period was relatively stable without significant variations in orbit or activity level \citep{Sekanina2003JGRE..108.8112S}.

The primary objective of our study is to constrain the potential "excavated volume" of the 81P depressions during its 30-year JFC phase (1974--2004) under the observational consideration of past activities. We studied the water and dust environment when 81P was inside the water snowline ($r_{\rm{h}} \lesssim 3\ \rm{au}$) based on previous ground-based observations. Given the inherent limitations stemming from incomplete in situ data on 81P activities, we addressed this gap through several models of water and dust with appropriate assumptions. We then compared the results with the observed 81P depression sizes and volumes. This comparison enables us to constrain the upper-limit contribution of 81P activity to the overall degree of depression enlargement.

In Sect. \ref{sec_VolumeOfDepressions}, we describe the methods and results to estimate the volumes and sizes of 81P depressions. Then, we present water (Sect. \ref{sec_WaterEnv}) and dust (Sect. \ref{sec_DustEnv}) environment of 81P to estimate the total excavated volume during its 30-year orbit, and compare the volume with that of depressions. We discuss the validity, reliability, and novelty of the results in Sect. \ref{sec_discussion}. We summarize this paper in Sect. \ref{sec_conclusion}.

\section{Volume of depressions}\label{sec_VolumeOfDepressions}

This section describes the methods and results for the depression volume calculation. First, we describe the method used to estimate depression volumes using Stardust images (Sects. \ref{subsec_ShapeModel}--\ref{subsec_DerivationOfDepressionVolume}). Then, we report our results of the depression volume estimates (Sect. \ref{subsec_Result01}).

\subsection{Shape model}\label{subsec_ShapeModel}

Images taken by the Stardust onboard camera Navcam \citep{Newburn2003JGRE..108.8116N} revealed approximately half of the 81P surface by capturing solar phase angle coverage ranging from $-72\degree$ to $+103\degree$. While these images revealed a predominant presence of quasicircular depressions, the shady region on the opposite side remains unexplored \citep{Brownlee2004Sci...304.1764B}. In this study, we use the terms "explored" and "unexplored" by Stardust to distinguish the two regions. The overall three-dimensional configuration of 81P closely conforms to a triaxial ellipsoid \citep{Duxbury2004JGRE..10912S02D}, with semiaxes measuring $2.607\ \rm{km}\ \times\ 2.002\ \rm{km}\ \times\ 1.350\ \rm{km}$ \citep{Kirk2005LPI....36.2244K}. The highest resolution achievable from the Navcam images was $14\ \rm{m\ pixel^{-1}}$ \citep{Brownlee2004Sci...304.1764B}. \citet{farnham2005shape} constructed a shape model based on these observed images. The model comprises a total of 17 513 facets, and the unexplored side is artificially extrapolated with the fit ellipsoid (Fig. \ref{fig_shapemodel01}a).

\subsection{Derivation of the depression volume} \label{subsec_DerivationOfDepressionVolume}

\begin{figure*}
\centering
\includegraphics[width=17cm]{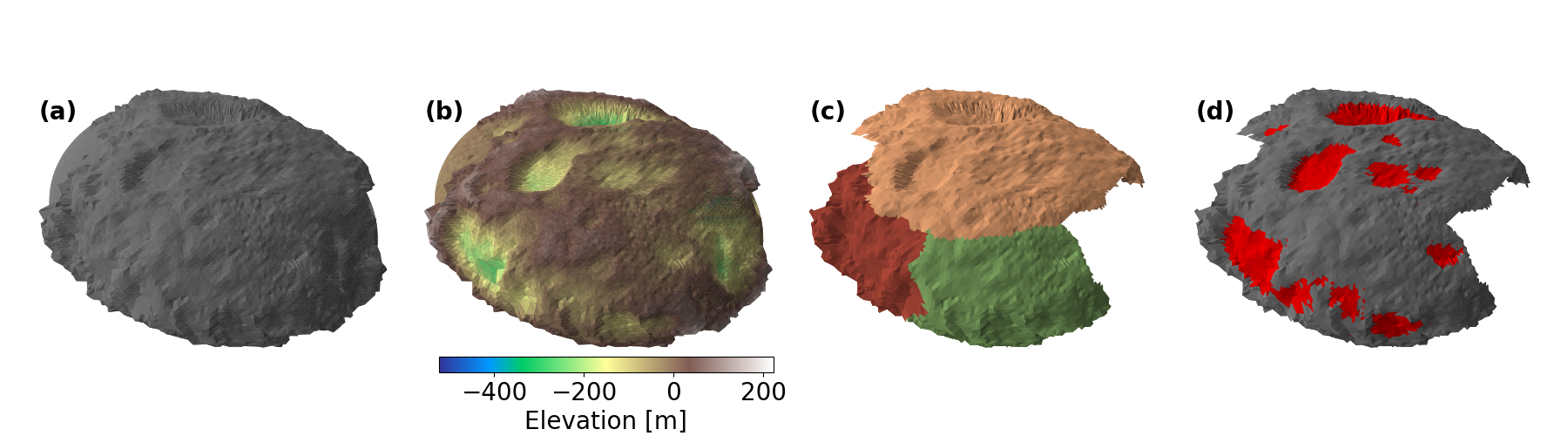}
\caption{\textbf{(a)} Shape model of 81P used in this study. The unexplored side is extrapolated to the fit ellipsoid.
\textbf{(b)} Elevation map of 81P.
\textbf{(c)} Three defined regions of the explored side: Left Foot \& Right Foot (orange), Rahe (red), and Hemenway (green).
\textbf{(d)} Depressions extracted by the edge detection technique (same as the outermost contours in Fig. \ref{fig_shapemodel02}).}
\label{fig_shapemodel01}
\end{figure*}

In this subsection, we outline the methodology for estimating the total volume of the 81P depressions. The term "volume" may carry ambiguity because these depressions are not entirely closed systems but are connected to neighboring topographies. We operationally define the volume as the surface "degree of the cavity," delineated by the deviation from the smooth zero-level potential. This definition aligns with the conventional framework employed to express the volume of impact craters \citep{Melosh1980AREPS...8...65M}. Importantly, these depressions are presumed to have been formed through excavation processes, enabling us to gauge the extent of surface excavation on 81P through this metric.

We determined the relative elevation of the facets by measuring the shortest distance between the point of each facet and the fit ellipsoid (Fig. \ref{fig_shapemodel01}b). We then used an edge detection technique to detect depressions and delineate borders. The Canny edge detector, a widely employed two-dimensional edge detection operator, was utilized for this purpose \citep{canny1986computational}. In principle, this detector identifies sharp boundaries in two-dimensional images marked by rapid elevation changes. To apply the detector to the 81P shape model, we utilized two-dimensional projected elevation maps of the original shape model. To minimize the distortion resulting from the two-dimensional projection of the three-dimensional ellipsoid, we divided the explored side into three distinct regions: Left Foot \& Right Foot, Rahe, and Hemenway (Fig. \ref{fig_shapemodel01}c).

The detector not only delineated the closed rims of the depressions but also highlighted various characteristic linear and curved features, potentially arising from the surface roughness of the comet nucleus \citep{Brownlee2004Sci...304.1764B}. To match the detector output with only the depression borders, we visually determined that the negative elevation contours were well aligned with the detector output. This process yielded "reference" zero-level elevations for these three regions, which were $\rm{-50\ m}$ for Left Foot \& Right Foot, $\rm{-80\ m}$ for Rahe, and $\rm{-40\ m}$ for Hemenway. By regarding the facets beneath these contour levels as the depressions (Fig. \ref{fig_shapemodel01}d), we computed the volume of the $j$-th depression as follows:
\begin{equation}\label{eq_shapemodel01}
    V_j = \sum_{i} d_i A_i \cos{\theta_i}~,
\end{equation}
where $d_i$ is the relative depth of the $i$-th facet from the reference contour level, $A_i$ is the surface area of the facet, and $\theta_i$ is the angle between the normal direction of the facet and the direction from the facet to the shortest point of the reference ellipsoid. The associated error of depths was considered by choosing the different points (three nodes and centers) on the facets, potentially containing a few meters of uncertainty for each facet \citep{farnham2005shape}.

We also estimated the dimensions (depth and diameter) of the depressions. We defined the depth of the $j$-th depression ($d_j$) as the distance between the deepest facet in the cavity and the reference contour level. Defining the depression diameter ($D$) was more complicated due to the irregular depression morphology. We derived two types of diameters, $D_{\rm{min}}$ and $D_{\rm{max}}$, which were defined as the diameters of the maximum circle enclosed by the depression borders and the minimum circle enclosing the depression borders, respectively. These definitions complement the previous definition of connecting two selected points on the opposite side of the borders \citep{Kirk2005LPI....36.2244K}.

\subsection{Results}\label{subsec_Result01}

\begin{figure*}
\centering
\includegraphics[width=17cm]{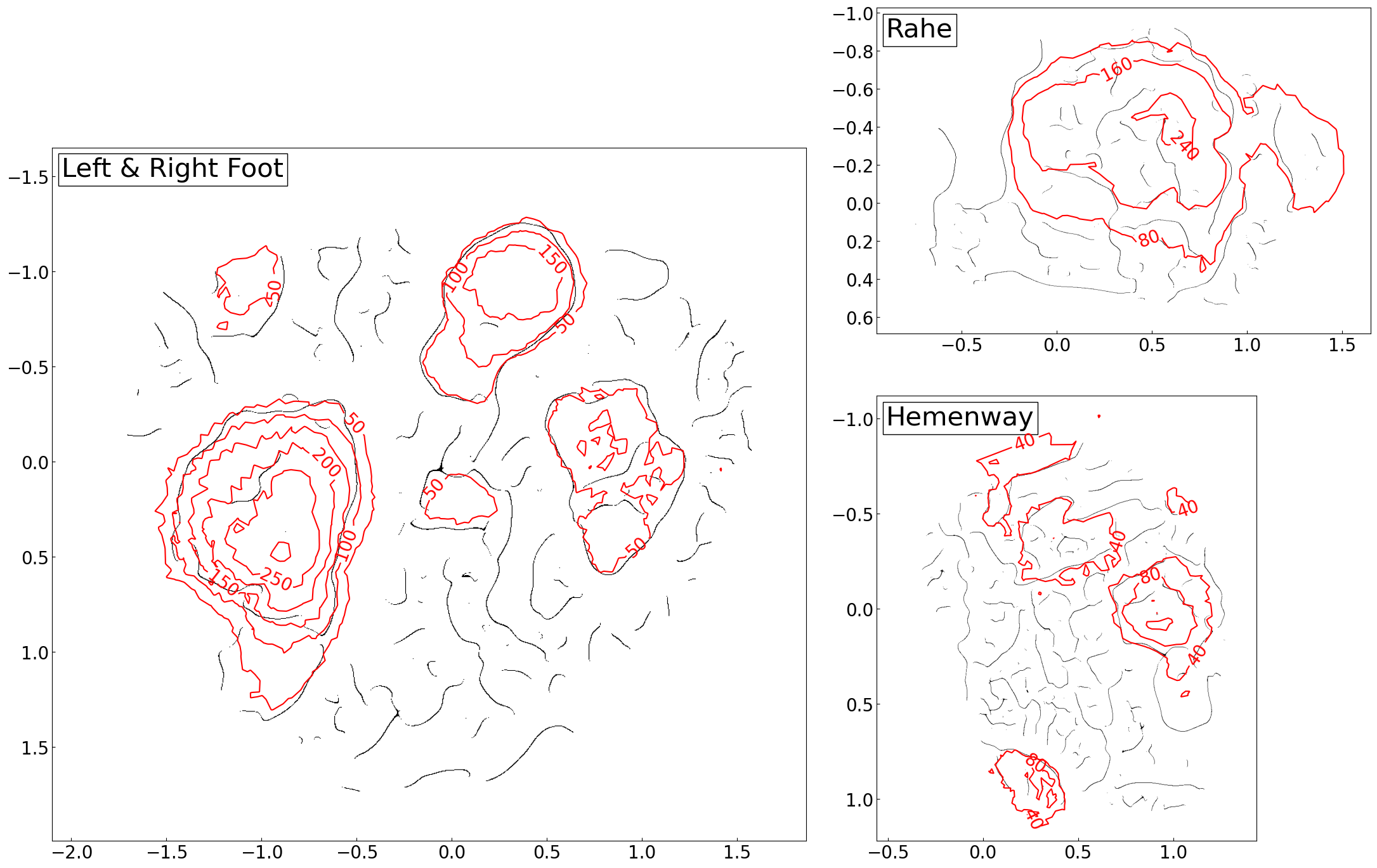}
\caption{Depressions were detected in three regions: Left Foot \& Right Foot, Rahe, and Hemenway. The black solid lines show the results of the edge detection technique. The red contours indicate the relative elevation labeled with depth (m). The outermost contours of each region represent the borders of the detected depressions. The x- and y-axes of each region are projected distances (km). In each plot, the horizontal and vertical axes indicate arbitrary Cartesian coordinates on the triaxial ellipsoid surface.}
\label{fig_shapemodel02}
\end{figure*}

\begin{table}

\caption{\label{table_result1}Depression sizes and volumes on 81P explored side.}
\centering
\setlength{\tabcolsep}{5pt}
\begin{tabular}{lcccccc}
\hline\hline
$\texttt{\#}$ & $d$ & $D_{\rm{min}}$ & $D_{\rm{max}}$ & Volume & Fraction & Region\tablefootmark{a} \\
 & [m] & [m] & [m] & [$\rm{km^3}$] & [\%] & \\
\hline
1 & 260 & 960 & 1620 & $.136^{+.012}_{-.022}$ & 46.7 & Right Foot \\
\hline
2 & 30 & 200 & 400 & .001 & 0.3 & \\
\hline
3 & 140 & 620 & 800 & $.035^{+.003}_{-.006}$ & 12.0 & Left Foot \\
4 & 50 & 360 & 460 &  &  & \\
\hline
5 & 30 & 200 & 400 & .001 & 0.3 & \\
\hline
6 & 65 & 500 & 600 & $.008^{+.001}_{-.002}$ & 2.8 & \\
7 & 45 & 300 & 420 &  &  & \\
8 & 30 & 140 & 220 &  &  & \\
\hline
9 & 180 & 1000 & 1280 & $.097^{+.008}_{-.016}$ & 33.3 & Rahe \\
10 & 70 & 300 & 660 &  &  & \\
\hline
11 & 45 & 200 & 400 & .001 & 0.4 & \\
12 & 30 & 100 & 240 &  &  & \\
\hline
13 & 40 & 360 & 580 & .002 & 0.8 & \\
\hline
14 & 90 & 440 & 660 & $.008^{+.001}_{-.002}$ & 2.7 & Hemenway\\
\hline
15 & 55 & 280 & 400 & .002 & 0.7 & \\
\hline
Total & & & & $.291^{+.025}_{-.048}$ & 100.0 & \\
\hline

\end{tabular}

\tablefoot{
\tablefoottext{a}{Refer to \citet{Brownlee2004Sci...304.1764B} for the nomenclature for each region.}
}

\end{table}

Using the edge detection technique for the 81P shape model, we detected a total of 15 depressions on the explored side. This detection is listed in Table \ref{table_result1} and shown in Fig. \ref{fig_shapemodel02}. Averages in depression volume were calculated by using the center of the facets, while the associated errors were calculated by using the three nodes of the facets. Because some depressions partially overlapped, we derived the total volumes for the overlapping depressions. In these cases, we distinguished each depression based on visual inspection.

We measured the depth ($d$) and diameters ($D_{\rm{min}}$ and $D_{\rm{max}}$) of each depression. The largest depressions in terms of diameter are located in Right Foot (\#1) and Rahe (\#9). These two depressions have $D>$ 1000 m. Due to the irregularity of the surface, some depressions show a factor of two difference between $D_{\rm{min}}$ and $D_{\rm{max}}$ (see Depressions \#2 and \#5 in Table \ref{table_result1}). The average depth-to-diameter ratios are $d/D_{\rm{min}}\sim0.2$ and $d/D_{\rm{max}}\sim0.12$, which are in line with previous estimates \citep[$d/D\sim0.21$;][]{Kirk2005LPI....36.2244K}.

We derived the total volume of these 15 depressions as $0.291^{+0.025}_{-0.048}\ \rm{km^3}$. The three largest depressions \citep[Right Foot, Left Foot, and Rahe;][]{Brownlee2004Sci...304.1764B} occupy most of this total volume. Left Foot and Rahe each likely consist of two depressions. Five depressions (Depressions \#1, \#3, \#4, \#9, and \#10) encompass $\sim$90 \% of the total depression volume. The Right Foot and Rahe regions, in particular, encompass almost $\sim$80 \% of the total depression volume and contain the two largest depressions (Depressions \#1 and \#9). Adopting the 81P nucleus bulk density \citep[$\rho_{\rm{bulk}}=600\ \rm{kg\ m^{-3}}$;][]{Davidsson2006Icar..180..224D}, the total volume of these depressions corresponds to a mass of $M_{\rm{depression}}=1.74^{+0.15}_{-0.29}\times 10^{11}\ \rm{kg}$.

\section{Water environments}\label{sec_WaterEnv}

In this section, we describe the water environments of 81P. We present our model to calculate the water sublimation in Sect. \ref{subsec_WaterSublimationModel} and the results in Sect. \ref{subsec_Result02}.

\subsection{Water sublimation model}\label{subsec_WaterSublimationModel}

In this subsection, we outline our 81P water sublimation model, which combines numerical simulation and observational constraints. The model enables us to acquire the full-phase water production curve throughout the orbit. The comet water production trend is inherently first-order dependent on the illuminated cross-section, and this dependency can result in asymmetric production before and after perihelion passage, particularly when the rotational pole is oblique \citep{Marshall2019A&A...623A.120M}. For this reason, we estimated the water sublimation environment, taking into account its known orbit, shape model (Sect. \ref{subsec_ShapeModel}), and spin axis orientation $(I, \it{\Phi}) = \rm{(55\degree, 150\degree)}$, where $I$ is the obliquity and $\it{\Phi}$ the longitude of the pole orientation \citep{Sekanina2004Sci...304.1769S}. Since the orbit remained stable for 30 years after the encounter with Jupiter in 1974, we did not consider the small variation in the orbit \citep{Sekanina2003JGRE..108.8112S}. Additionally, we did not consider the nongravitational effect of the 81P spin state due to its small change \citep{Gutierrez2007Icar..191..651G}.

We simulated the water sublimation environment following the frameworks established in previous works \citep[e.g.,][]{Steckloff2015Icar..258..430S, Marshall2019A&A...623A.120M}. We postulated a cometary surface where sublimation activity occurs instantaneously (i.e., zero thermal inertia) and is fully homogeneous across the nucleus surface (i.e., pure water ice surface). Under these assumptions, the energy balance among solar radiation, thermal emission, and water ice sublimation on the nucleus surface is given by
\begin{equation}
\label{eq_water01}
    \frac{S_{\sun}\left(1-A_{\rm{H}}\right)}{r_{\rm{h}}^2}\cos{\theta_i}=\varepsilon \sigma T_i^4+Z(T_i)L_{\rm{ice}}~,
\end{equation}
where the subscript $i$ indicates the $i$-th facet of the shape model, $S_{\sun}=1367\ \rm{W\ m^{-2}}$ is the solar constant equivalent to the solar flux at 1 au, $A_{\rm{H}}=0.01$ is the bolometric Bond albedo \citep{Lamy2007SSRv..128...23L}, $r_{\rm{h}}$ is the heliocentric distance in units of au (practically dimensionless in the equation), $\theta_i$ is the solar incidence angle of the facet, $\varepsilon=0.9$ is the emissivity, $\sigma$ is the Stefan--Boltzmann constant, $T_i$ is the sublimation temperature of the facet, and $L_{\rm{ice}}=2.6\times10^6\ \rm{J\ kg^{-1}}$ is the latent heat of water ice.

In Eq. (\ref{eq_water01}), $Z(T)\ (\rm{kg\ s^{-1}\ m^{-2}})$ is the temperature-dependent water sublimation rate. It is given by Hertz--Knudsen equation \citep{Steiner1991JGR....9618897S}:
\begin{equation}
\label{eq_water02}
    Z(T)=\frac{p_{\rm{evp}}(T)}{\sqrt{2\pi m_{\rm{H_{2}O}}k_{\rm{B}}T}} ~,
\end{equation}
where ${p_{\rm{evp}}(T)}$ is the evaporation gas pressure obtained by solving the Clausius--Clapeyron equations \citep{Keller1989A&A...226L...9K}, $m_{\rm{H_{2}O}}$ is the water molecular mass, and $k_{\rm{B}}$ is the Stefan--Boltzmann constant. Combining Eqs. (\ref{eq_water01}) and (\ref{eq_water02}) enables us to calculate the equilibrium sublimation temperature and the corresponding sublimation rate on the facets at a given solar incident flux. Then, we estimated the total water production rate ($Q_{\rm{H_{2}O}}$; $\rm{molecule\ s^{-1}}$) on the entire nucleus surface at a given heliocentric distance by the following equation:
\begin{equation}
\label{eq_water03}
    m_{\rm{H_{2}O}}{Q_{\rm{H_{2}O}}}(r_{\rm{h}})=\sum_{i}Z_{i}(r_{\rm{h}})A_{i} ~.
\end{equation}

We computed the full-phase water production rates (Eq. (\ref{eq_water03})) for 200 evenly divided true anomalies. The water sublimation rate for every orbital step was averaged daily by rotating the subsolar point over one comet day at 25 uniform intervals. For each computation, we considered only the solar-facing facets ($\theta_i<90\degree$). We did not consider the change in the heliocentric distance during one comet rotation ($\Delta r_{\rm{h}}\sim10^{-4}\ \rm{au}$ per comet rotation). We ignored the effects of shadowing and self-heating by other facets.

We confined our simulations to $r_{\rm{h}}<3\ \rm{au}$ not only because the water sublimation drops significantly around $r_{\rm{h}}=3\ \rm{au}$ but also because the simplified approach described above loses validity beyond this distance. Beyond this line ($r_{\rm{h}}>3\ \rm{au}$), the role of heat conduction becomes nonnegligible \citep{Steckloff2015Icar..258..430S}, and the primary sublimation species shift from water to other supervolatiles, such as $\rm{CO}$ or $\rm{CO_2}$ \citep{Davidsson2022MNRAS.509.3065D}. Neither of these factors were incorporated into our formulations. While recognizing the potential limitations of these simplifications, we justified our approach as suitable for our specific objective, considering that water sublimation-driven erosion is the predominant mechanism for surface mass loss \citep{Kossacki2018Icar..305....1K}.

Nevertheless, the areal water production rates are highly overestimated under the assumptions above---zero thermal inertia and a pure water ice surface---because a large fraction of the surface could be covered with an inert dust mantle. In pursuit of a more realistic representation, we modified the second assumption by considering that the surface comprises well-mixed pure water ice and refractories. This involved introducing a scaling parameter $f$ into the right-hand side of Eq. (\ref{eq_water03}), that is,
\begin{equation}
\label{eq_water04}
    m_{\rm{H_{2}O}}Q_{\rm{H_{2}O}}(r_{\rm{h}})=f\sum_{i}Z_{i}(r_{\rm{h}})A_{i}\\(0\leq f\leq1)~,
\end{equation}
where $f$ is often referred to as an "effective active fraction" and is interpreted as the fraction of pure water ice on the surface; it is used to scale the idealized free water sublimation rate to the real rate \citep{Keller1989A&A...226L...9K}. The value of $f$ does not convey a physical meaning for characterizing real cometary surfaces, as the expected sublimation process is more intricate (e.g., involving sublimation from subsurface ice).

We constrained the possible range of $f$ by considering the observed water production rate data and their standard errors. We used the water production rates during the two perihelion passages in 1997 and 2010, when extensive observations were conducted due to favorable observation conditions \citep{Fink1999Icar..141..331F}. These data sets are summarized in Table \ref{table_water1}. Since single-night water production measurements often have large errors, we binned the data points at 50-day intervals and used the average values within each bin. Referring to the variance-defined weights, we calculated the weighted mean water production rate in each interval as
\begin{equation}\label{eq_water05}
    Q_{50}=\frac{\sum_{i}^N w_i Q_i}{\sum_{i}^N w_i} ~,
\end{equation}
where $N$ is the number of water production data points within each bin, $Q_i$ is the individual measurement of the water production rate within the interval, $w_i=1/\sigma_i^2$ is the weight defined by the variance, and $\sigma_i$ is the standard error of the water production rate from the individual measurements. Here, the subscript $\rm{H_2O}$ is omitted for simplicity. Similarly, the weighted standard error in each interval is calculated as
\begin{equation}\label{eq_water06}
    \sigma_{50}= \sqrt{\frac{\sum_{i}^N w_i (Q_i - Q_{50})^2}{\frac{N}{N-1} \sum_{i}^N w_i}} ~.
\end{equation}
The water production rates calculated using Eqs. (\ref{eq_water05}) and (\ref{eq_water06}) are included in Table \ref{table_water1}. We set the lower and upper limits of the effective active fraction ($f_{\rm{min}}$ and $f_{\rm{max}}$) according to the values enclosing all the binned data and their standard errors (see Fig. \ref{fig_water01}).

Specifically, the effective active fraction $f$ is a parameter that varies with the heliocentric distance and with the nucleus coordinates if the surface is heterogeneous. In general, $f$ tends to decrease at a larger heliocentric distance \citep{Marschall2020FrP.....8..227M} and shows large surface heterogeneity \citep{Attree2019A&A...630A..18A}. We did not rigorously include this effect in our study.

\subsection{Results}\label{subsec_Result02}

\begin{figure}
\resizebox{\hsize}{!}{\includegraphics{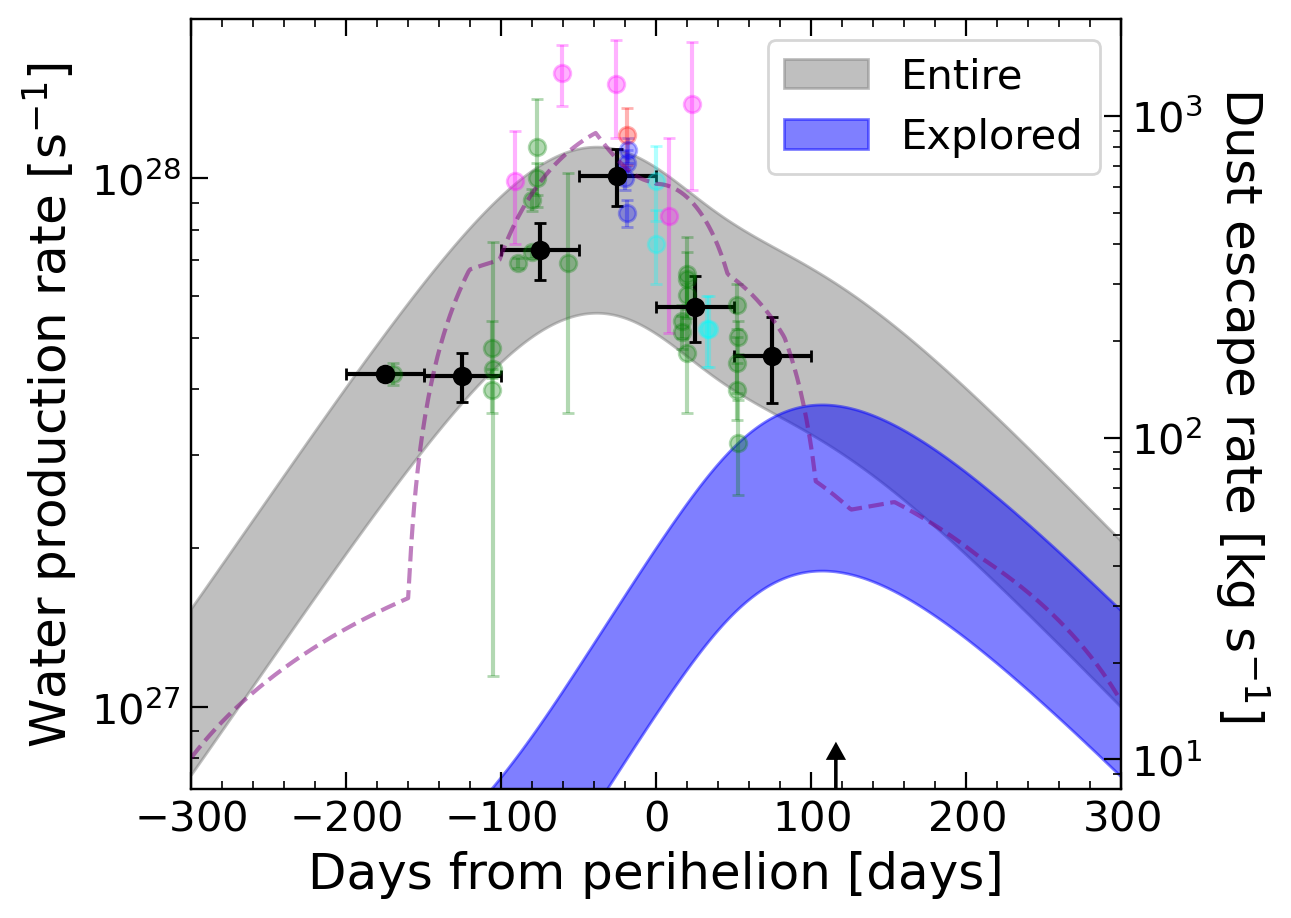}}
\caption{Seasonal water production rate of 81P scaled with observational data. The red, green, blue, magenta, and cyan points show the observational data from previous studies, and the black-filled circles are the average data within the 50-day bin (Table \ref{table_water1}). The gray and blue-filled regions represent the possible ranges of the water production rates from the entire surface (gray) and from only the explored side (blue), respectively, which is scaled by the average data. The dotted line shows the dust escape rates adopted from \citet{Pozuelos2014A&A...571A..64P}. The black arrow indicates the day on which Stardust encountered 81P.} 
\label{fig_water01}
\end{figure}

\begin{figure*}
\centering
\includegraphics[width=17cm]{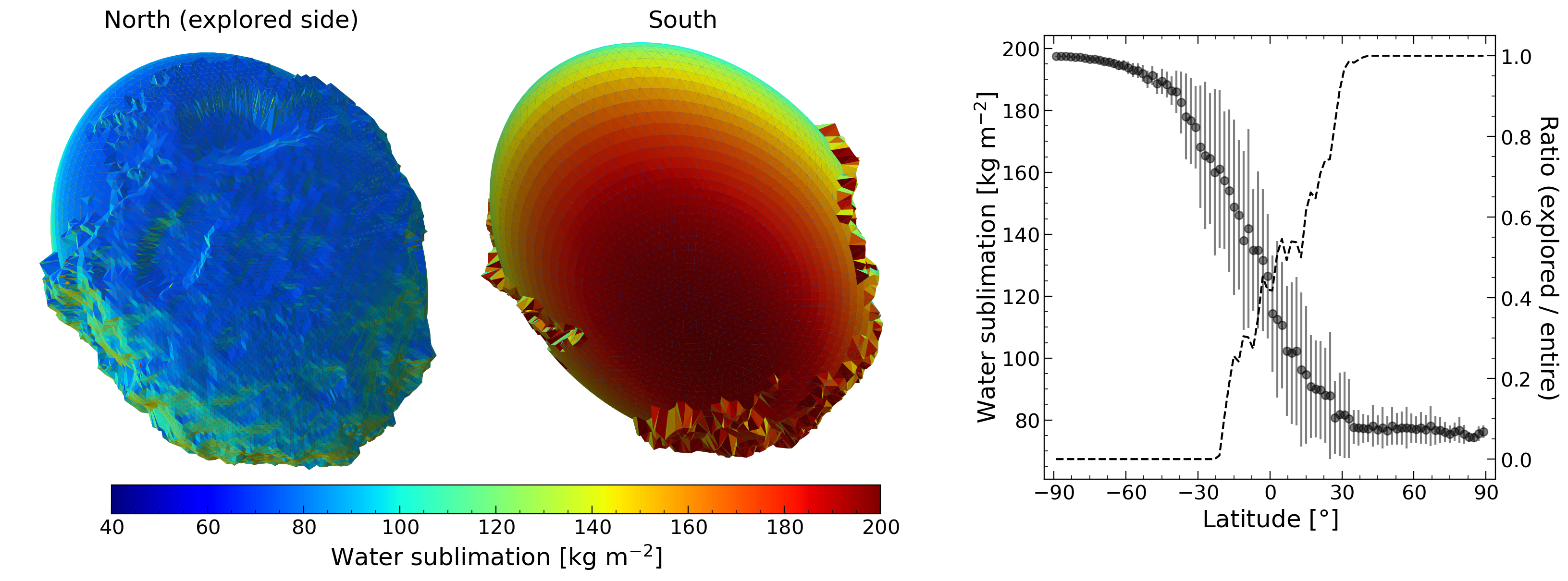}
\caption{(left) Orbit-integrated water sublimation of 81P facets. The left and right figures preferentially show the explored northern and the unexplored southern side, respectively. (right) Latitudinal-dependent water production. The filled circles and the error bars show average sublimation rates and standard deviations of the facets inside the latitudinal bin. The dotted line shows the fraction of the explored side with respect to the entire surface inside the latitudinal bin.}
\label{fig_water02}
\end{figure*}

We simulated seasonal water production rates under simplified conditions: a homogeneous surface with zero thermal inertia. Here, we considered the orbital motion, the shape model, and the measured spin orientation. We averaged the individual observational data with 50-day intervals and scaled the simulated water production curve into the binned data. Here, the range of the scaling factor ($f$) was determined by encompassing the whole binned data and their errors. This criterion produces the same order of water activity trends with observational data while mitigating the large scatters of individual observations.

Fig. \ref{fig_water01} (gray-filled region) presents the resultant seasonal water production rate. It reveals asymmetric water production with respect to the perihelion, peaking approximately 40 days before the perihelion passage. This computational result is in line with the observed peak of the 81P water production rate \citep[40--60 days before the perihelion;][]{deValBorro2010A&A...521L..50D}. The overall trend of the water production roughly follows the dust escape rate (dotted line), which was reported by \citet{Pozuelos2014A&A...571A..64P}. The peak of dust escape rate also occurs 40 days before the perihelion.

We find that $f$ is in the range of 0.08 to 0.16 (the lower and upper ends of filled regions in Fig. \ref{fig_water01} correspond to $f=0.08$ and $0.16$, respectively). We derive the peak water production rates for the minimum and maximum cases as $Q_{\rm{H_2O}} \sim 5.6 \times 10^{27}\ \rm{molecule\ s^{-1}}$ and $Q_{\rm{H_2O}}\sim11.5 \times 10^{27}\ \rm{molecule\ s^{-1}}$, respectively, and the total water loss during one comet orbit as $4.6$--$9.5 \times 10^9\ \rm{kg}$. When we limit the calculation to the explored side (blue-filled region), we find that the surface lost $1.3$--$2.7 \times 10^9\ \rm{kg}$ of water during one comet orbit (29 \% of the total water loss).

The orbit-integrated water production in the southern hemisphere is two to three times greater than in the northern hemisphere (Fig. \ref{fig_water02}). The tilted pole orientation to the orbital plane can explain this latitudinal asymmetry. The illuminated area and subsequent water production on the inbound trajectory are twice as large as those on the outbound trajectory due to the tilted pole \citep{deValBorro2010A&A...521L..50D}. During the period around the perihelion passage, the subsolar latitude gradually moves from the southern to the northern side.

As a result of asymmetric water production, the explored side (mostly the northern hemisphere) contributes to a total activity of $\sim$30 \% during one cometary orbit around the Sun, while the same area consists of $\sim$40 \% of the entire surface ($22$ of $56\ \rm{km^2}$). The Stardust spacecraft flew by 81P approximately 116 days after its perihelion passage (black arrow in Fig. \ref{fig_water01}), making it possible to observe the northern hemisphere. Our calculation result for the water production rate suggests that this comet's activity had already been reduced by a factor of two with respect to its peak and that it was observed in the less active northern hemisphere (opposite the active southern hemisphere).

The explored side in Fig. \ref{fig_water02} shows a heterogeneous water sublimation among the facets, potentially depending on the slopes. This may accelerate the lateral expansion of the depressions rather than deepening the bottoms, as observed in 67P \citep{Attree2018A&A...610A..76A, Benseguane2022A&A...668A.132B, GuilbertLepoutre2023PSJ.....4..220G}. We further discuss the heterogeneous erosion of the explored side in Sect. \ref{subsec_LocationOfMassExcavation}.

\section{Dust environments}\label{sec_DustEnv}

In this section, we describe the dust environments of 81P. We present our dust ejection model in Sect. \ref{subsec_DustEjectionModel} and the results in Sect. \ref{subsec_Result03}. Finally, we summarize the results of depressions, water sublimation, and dust ejection of 81P and derive the depression excavation rate in Sect. \ref{subsec_ExcavationRate}.

\subsection{Dust ejection model}\label{subsec_DustEjectionModel}

\begin{figure*}
\centering
\includegraphics[width=17cm]{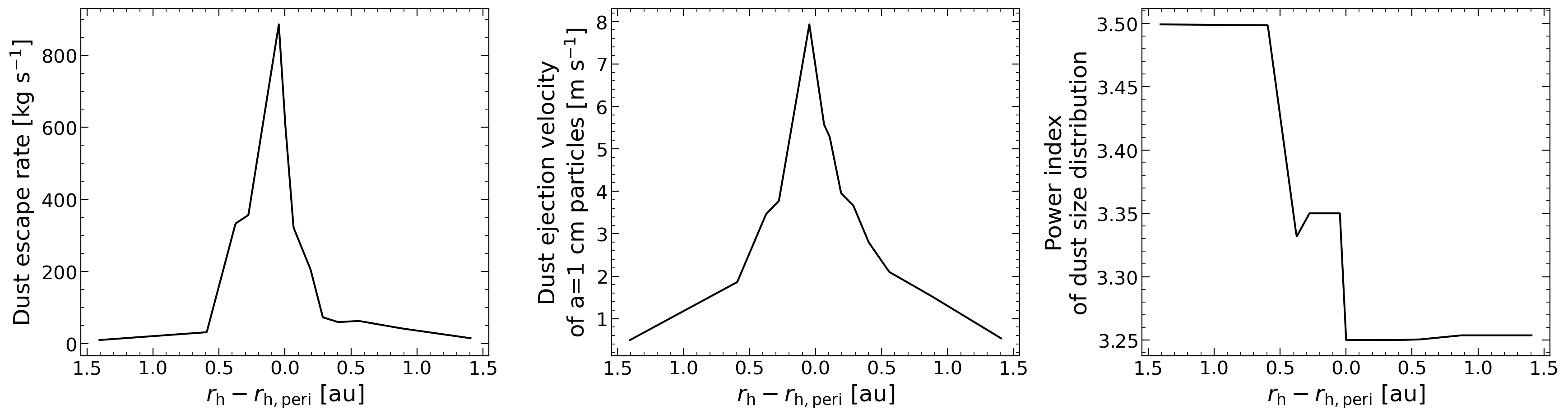}
\caption{Heliocentric dependencies of (left) the dust ejection rate, (center) the dust velocity of $a_0=1\ \rm{cm}$-sized particles at a cometocentric distance of $R=20R_{\rm{N}}$, and (right) the power index of the dust size distribution. The plots are adopted from Fig. 3 in \citet{Pozuelos2014A&A...571A..64P}, while we excluded the data relevant to the two small outburst events observed in the 2010 apparition.
}
\label{fig_dust01}
\end{figure*}

\begin{figure*}
\includegraphics[width=17cm]{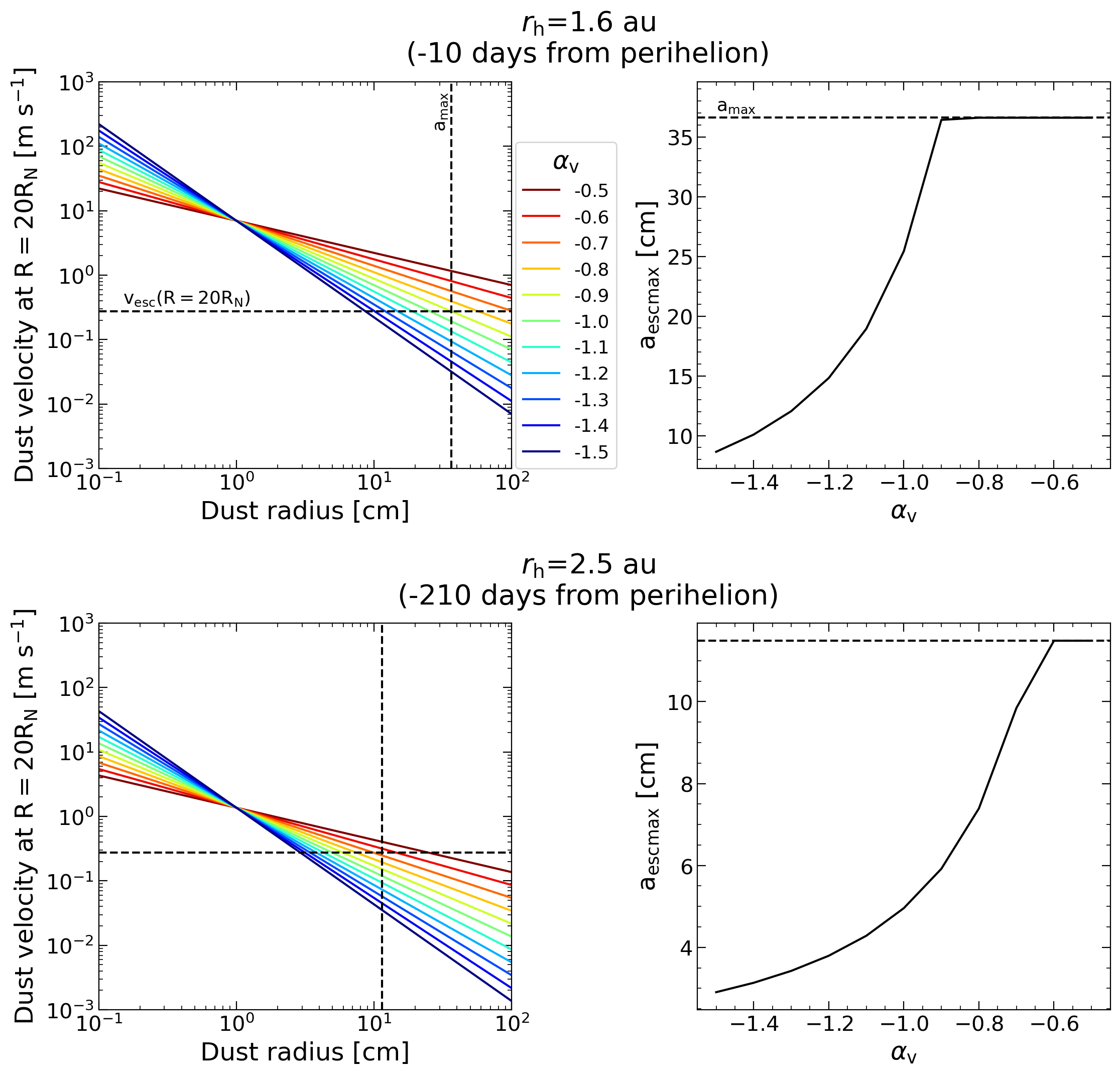}
\caption{Dust velocity ($v_{\rm{d}}$) at $R=20R_{\rm{N}}$ and maximum escape radius ($a_{\rm{escmax}}$) in our dust ejection model. The upper and lower panels represent examples at small and large heliocentric distances, respectively. The left column shows the size-dependent dust velocity at a cometocentric distance of $R=20R_{\rm{N}}$. The horizontal dotted lines indicate the escape velocity ($v_{\rm{esc}}$) at the cometocentric distance. The vertical dotted lines indicate the maximum liftable radius ($a_{\rm{max}}$) at the heliocentric distance. The intersection points with horizontal dotted and colored lines correspond to the maximum escape radius at the given $\alpha_{\rm{v}}$ values. The right column shows the determined maximum escape radius as a function of $\alpha_{\rm{v}}$. The horizontal dotted lines represent the maximum liftable radius, showing the upper limit of the maximum escape radius at given heliocentric distances. }
\label{fig_dust02}
\end{figure*}

In this subsection, we outline our model for estimating the amount of dust ejected from the nucleus surface. Dust particles can be ejected from the nucleus surface primarily due to the gas lifting force generated by the expansion of sublimating volatiles \citep{Grun1989AdSpR...9c.133G}. In a general examination of the interplay among the lifting force, nucleus gravity, and dust particle properties, we identify three distinct cases of the dust ejection process:
\begin{enumerate}[label=(\roman*)]
    \item Dust particles that are ejected from the surface and eventually escape from the comet Hill sphere ($\dot{M}_{\rm{escape}}$),
    \item Dust particles that are ejected from the surface but subsequently fall back to the surface ($\dot{M}_{\rm{fallback}}$),
    \item Dust particles that remain on the surface without lifting. \label{item_dust3}
\end{enumerate}
Here, we define $\dot{M}_{\rm{escape}}$ ($\rm{kg\ s^{-1}}$) as the dust escape rate and $\dot{M}_{\rm{fallback}}$ ($\rm{kg\ s^{-1}}$) as the dust fallback rate. Although it is not critical to consider the dust particles in case (iii) in this section, they should be taken into account to fully describe the nucleus surface environment.

Given the water production environment that determines the lifting force, the differences among these three cases depend solely on the dust particle mass (or size) if we assume homogeneous and spherical dust particles ($m_{\rm{d}}=4\pi a^3 \rho_{\rm{d}}/3$, where $a$ and $\rho_{\rm{d}}$ denote the dust radius and the bulk mass density, respectively). Here, we set $\rho_{\rm{d}}=600\ \rm{kg\ m^{-3}}$ for dust particles from 81P \citep{Niimi2012ApJ...744...18N}.

Assuming that the size distribution of the dust production follows a single-power law, the number of dust particles ejected in a unit of time within the size range of $a_{\rm{min}}<a<a_{\rm{max}}$ is expressed as
\begin{equation}\label{eq_dust01}
    n_{\rm{ej}}(a)=k_{\rm{p}}\,\left(\frac{a}{a_0}\right)^{-b_{\rm{d}}}\\(a_{\rm{min}}<a<a_{\rm{max}}) ~,
\end{equation}
where $k_{\rm{p}}\ (\rm{s^{-1}})$ is a normalization parameter, $b_{\rm{d}}$ is the power index of the size distribution, and $a_0=1\ \rm{cm}$ is the reference dust radius. We fixed the minimum radius of the dust particles to $a_{\rm{min}}=1\ \rm{\mu m}$ \citep{Pozuelos2014A&A...571A..64P}. $a_{\rm{max}}$ represents the "maximum liftable radius," which is determined by the balance between the gas drag force and the nucleus gravity at the surface. Assuming adiabatic spherical expanding gas, an analytical solution for $a_{\rm{max}}$ can be expressed as \citep{Zakharov2018Icar..312..121Z}
\begin{equation}\label{eq_dust02}
    a_{\rm{max}}(r_{\rm{h}}) = \frac{3 m_{\rm{H_2O}} Z(r_{\rm{h}}) R^2_{\rm{N}} C_{\rm{D}} \sqrt{k_{\rm{B}} T^\ast \gamma / m_{\rm{H_2O}}}}{8 \rho_{\rm{d}} M_{\rm{N}} G} ~, 
\end{equation} 
where $\gamma = 4/3$ is the specific heat ratio of water, $G$ is the gravitational constant, $R_{\rm{N}}=2\ \rm{km}$ is the effective nucleus radius, and $M_{\rm{N}}=2.3\times10^{13}\ \rm{kg}$ is the nucleus mass \citep[$\rho_{\rm{bulk}}=600\ \rm{kg\ m^{-3}}$;][]{Davidsson2006Icar..180..224D}. $T^\ast$ is the gas temperature on the sonic surface, which is related to the sublimation temperature ($T$ in Eq. (\ref{eq_water01})) by $T^\ast=0.8T$ \citep{lukianov2000stationary}. $C_{\rm{D}}$ is the gas drag coefficient; for a spherical dust particle, it can be expressed as follows \citep{bird1994molecular}:
\begin{equation}\label{eq_dust03}
    C_{\rm{D}}(s) = \frac{2s^2+1}{s^3\sqrt{\pi}}\exp{(-s^2)} + \frac{4s^4+4s^2-1}{2s^4}\rm{erf}(s) + \frac{2\sqrt{\pi}}{3s} ~,
\end{equation} 
where $s$ is the speed ratio between the expanding gas and the gas thermal velocity. The adiabatic solution yields $s=\sqrt{\gamma/2}$ at the surface. Notably, $a_{\rm{max}}$ depends on the local water sublimation rate ($Z$) and therefore depends on the heliocentric distance along the comet orbit.

The sum of $\dot{M}_{\rm{escape}}$ and $\dot{M}_{\rm{fallback}}$ is calculated using $n_{\rm{ej}}$ as
\begin{equation}\label{eq_dust04}
    \dot{M}_{\rm{escape}} + \dot{M}_{\rm{fallback}} = \int_{a_{\rm{min}}}^{a_{\rm{max}}} m_{\rm{d}}(a)\left|\frac{dn_{\rm{ej}}(a)}{da}\right|\,da~.
\end{equation}
Eq. (\ref{eq_dust04}) is obtained based on mass flux conservation, indicating that dust either escapes or falls back once ejected. We did not account for the disintegration of particles, which could alter the dust size distribution over time \citep{Davidsson2024MNRAS.527..112D}. We separated Eq. (\ref{eq_dust04}) as follows:
\begin{equation}\label{eq_dust05}
\begin{split}
    \dot{M}_{\rm{escape}}&=\int_{a_{\rm{min}}}^{a_{\rm{escmax}}} m_{\rm{d}}(a)\left|\frac{dn_{\rm{ej}}(a)}{da}\right|\,da\\
    &= \frac{4\pi}{3} \frac{b_{\rm{d}}}{3-b_{\rm{d}}} \rho_{\rm{d}} k_{\rm{p}} \left[ \left(\frac{a_{\rm{escmax}}}{a_0}\right)^{-b_{\rm{d}}} a_{\rm{escmax}}^3 - \left(\frac{a_{\rm{min}}}{a_0}\right)^{-b_{\rm{d}}} a_{\rm{min}}^3\right]~,
\end{split}
\end{equation}
and
\begin{equation}\label{eq_dust06}
\begin{split}
    \dot{M}_{\rm{fallback}}&=\int_{a_{\rm{escmax}}}^{a_{\rm{max}}} m_{\rm{d}}(a)\left|\frac{dn_{\rm{ej}}(a)}{da}\right|\,da\\
    &= \frac{4\pi}{3} \frac{b_{\rm{d}}}{3-b_{\rm{d}}} \rho_{\rm{d}} k_{\rm{p}} \left[ \left(\frac{a_{\rm{max}}}{a_0}\right)^{-b_{\rm{d}}} a_{\rm{max}}^3 - \left(\frac{a_{\rm{escmax}}}{a_0}\right)^{-b_{\rm{d}}} a_{\rm{escmax}}^3 \right]~,
\end{split}
\end{equation}
where $a_{\rm{escmax}}$ represents the "maximum escape radius," which distinguishes between escaping and falling back dust particles. Accordingly, $a_{\rm{escmax}}$ is a key parameter for determining $\dot{M}_{\rm{escape}}$ and $\dot{M}_{\rm{fallback}}$. Here, $\dot{M}_{\rm{escape}}$ is a quantity that can be measured from ground-based observations of dust comas and dust tails. In fact, \citet{Pozuelos2014A&A...571A..64P} conducted intensive observations of the dust coma and tail of 81P and derived the time-dependent $\dot{M}_{\rm{escape}}$. In contrast, $\dot{M}_{\rm{fallback}}$ cannot be determined directly from ground-based observations but can be predicted using a possible range of $a_{\rm{escmax}}$, as described below.

To determine $a_{\rm{escmax}}$, we utilized the 81P dust ejection model constrained by \citet{Pozuelos2014A&A...571A..64P}. During the 2010 apparition, the authors conducted a comprehensive study on the evolution of the optical coma brightness and the dust tail morphology of 81P across a broad heliocentric distance range both before ($r_{\rm{h}}<2.7\ \rm{au}$) and after ($r_{\rm{h}}<3.2\ \rm{au}$) the perihelion passage. \citet{Pozuelos2014A&A...571A..64P} used $\sim$300 ground-based observation data points obtained through 2010 apparition, which were mostly observed by amateur astronomical association \it Cometas-Obs \rm (to find the complete dataset, refer to their Fig. 4). The authors fitted the data with Monte Carlo dust tail modeling \citep{Moreno2009ApJS..183...33M} to analyze the dust tail morphology in terms of the initial dust ejection parameters \citep[see][for additional information on the usage of the model]{Pozuelos2014A&A...568A...3P}. This approach enabled them to constrain the seasonal variations in the dust escape rate ($\dot{M}_{\rm{escape}}$), the dust velocity along the normal direction ($v_{\rm{d}}$) of $a_0=1\ \rm{cm}$ particles at $R=20R_{\rm{N}}$, where $R$ is the cometocentric distance, and the power index of the dust size distribution ($b_{\rm{d}}$). Fig. \ref{fig_dust01} presents a reproduction of their dust modeling, adopted from Fig. 3 in \citet{Pozuelos2014A&A...571A..64P}. We excluded two minor transient outbursts that occurred in 2010 \citep{Pozuelos2014A&A...571A..64P} to characterize the general trend of the activity. The effects of these cometary outbursts on the depressions are discussed separately in Sect. \ref{subsubsec_Outburst}.

Following the works of \citet{Pozuelos2014A&A...571A..64P}, we considered dust velocity ($v_{\rm{d}}$) of $a_0=1\ \rm{cm}$ particles at $R=20R_{\rm{N}}$. At this distance, the gas drag force on the dust particles is deemed negligible, and $v_{\rm{d}}$ reaches the terminal velocity \citep{Zakharov2018Icar..312..121Z, Patzold2019MNRAS.483.2337P}. We further assumed that $v_{\rm{d}}$ can be determined by the dust size. Accordingly, the size-dependent $v_{\rm{d}}$ is expressed as
\begin{equation}\label{eq_dust07}
    v_{\rm{d}}(a,R=20R_{\rm{N}})=v_{\rm{d}}(a=a_0,R=20R_{\rm{N}})\left(\frac{a} {a_0}\right)^{-\alpha_{\rm{v}}} ~,
\end{equation}
where $\alpha_{\rm{v}}$ is the power index of the size-dependent velocity distribution at a cometocentric distance of $R=20R_{\rm{N}}$ (i.e., when the dust reaches the terminal velocity). \citet{wallis1982dusty} derived $\alpha_{\rm{v}}\sim0.5$ via an analytical approach. In-situ measurements obtained using GIADA onboard Rosetta yielded $\alpha_{\rm{v}}=0.96\pm0.54$ at $5R_{\rm{N}} \lesssim R \lesssim 15R_{\rm{N}}$ from 67P \citep{DellaCorte2015A&A...583A..13D}. Given the large uncertainty associated with these in situ measurements, we considered $\alpha_{\rm{v}}=1.0\pm0.5$.

Eq. (\ref{eq_dust07}) enables us to determine $a_{\rm{escmax}}$ by comparing $v_{\rm{d}}$ with the escape velocity $v_{\rm{esc}}(R)=\sqrt{2GM_{\rm{N}}/R}$ at $R=20R_{\rm{N}}$. Fig. \ref{fig_dust02} illustrates the examples of how the $a_{\rm{escmax}}$ values are determined in two cases at small and large heliocentric distances. In the left column, the escape velocity at $R=20R_{\rm{N}}$ (horizontal dotted line) is fixed to $v_{\rm{esc}}=0.23\ \rm{m\ s^{-1}}$, and the $a_{\rm{max}}$ (vertical dotted line) is determined by the water production rate (Eq. (\ref{eq_dust02})). Then, $a_{\rm{escmax}}$ is determined based on the intersection points between $v_{\rm{d}}$ and $v_{\rm{esc}}$ or $v_{\rm{d}}$ and $a_{\rm{max}}$. As a result, the particles escape if the dust velocity is smaller than the escape velocity ($v_{\rm{d}}<v_{\rm{esc}}$) and fall back if the dust velocity is larger ($v_{\rm{d}}>v_{\rm{esc}}$). As seen from the right column of Fig. \ref{fig_dust02}, $\alpha_{\rm{v}}$ is the key parameter for determining $a_{\rm{escmax}}$. Considering the full range of $\alpha_{\rm{v}}$ (i.e., $0.5\leq\alpha_{\rm{v}}\leq1.5$), there is a factor of three difference in $a_{\rm{escmax}}$ at the small heliocentric distance. This tendency decreases when the comet is far from the Sun because weak solar insolation at a large heliocentric distance narrows the gap between $v_{\rm{d}}(a_0=1\ \rm{cm})$ and $v_{\rm{esc}}$.

We did not incorporate additional forces, such as the solar radiative force or solar gravity, which are deemed negligible in this region in the short time period under consideration here \citep{Marschall2020FrP.....8..227M}. Consequently, we can simplify the analysis of dust motion in this regime by focusing solely on the nucleus gravity.

\subsection{Results}\label{subsec_Result03}

\begin{figure}
\resizebox{\hsize}{!}{\includegraphics{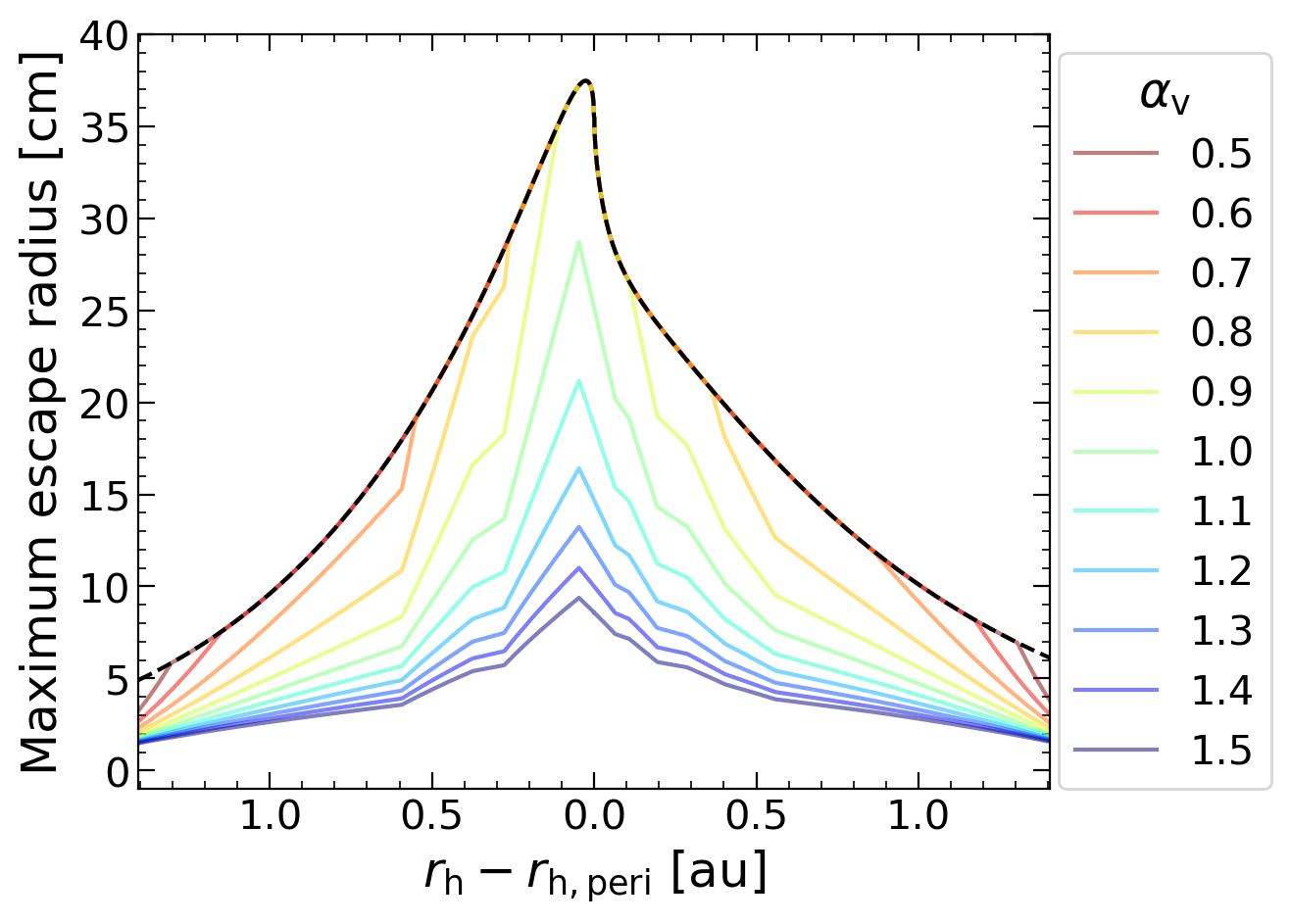}}
\caption{Maximum escape radius ($a_{\rm{escmax}}$) at different heliocentric distances and $\alpha_{\rm{v}}$ values. The dotted line represents the maximum liftable radius and indicates the upper limit of the maximum escape radius; thus, the colored lines cannot exceed this limit.}
\label{fig_dust03}
\end{figure}

\begin{figure}
\resizebox{\hsize}{!}{\includegraphics{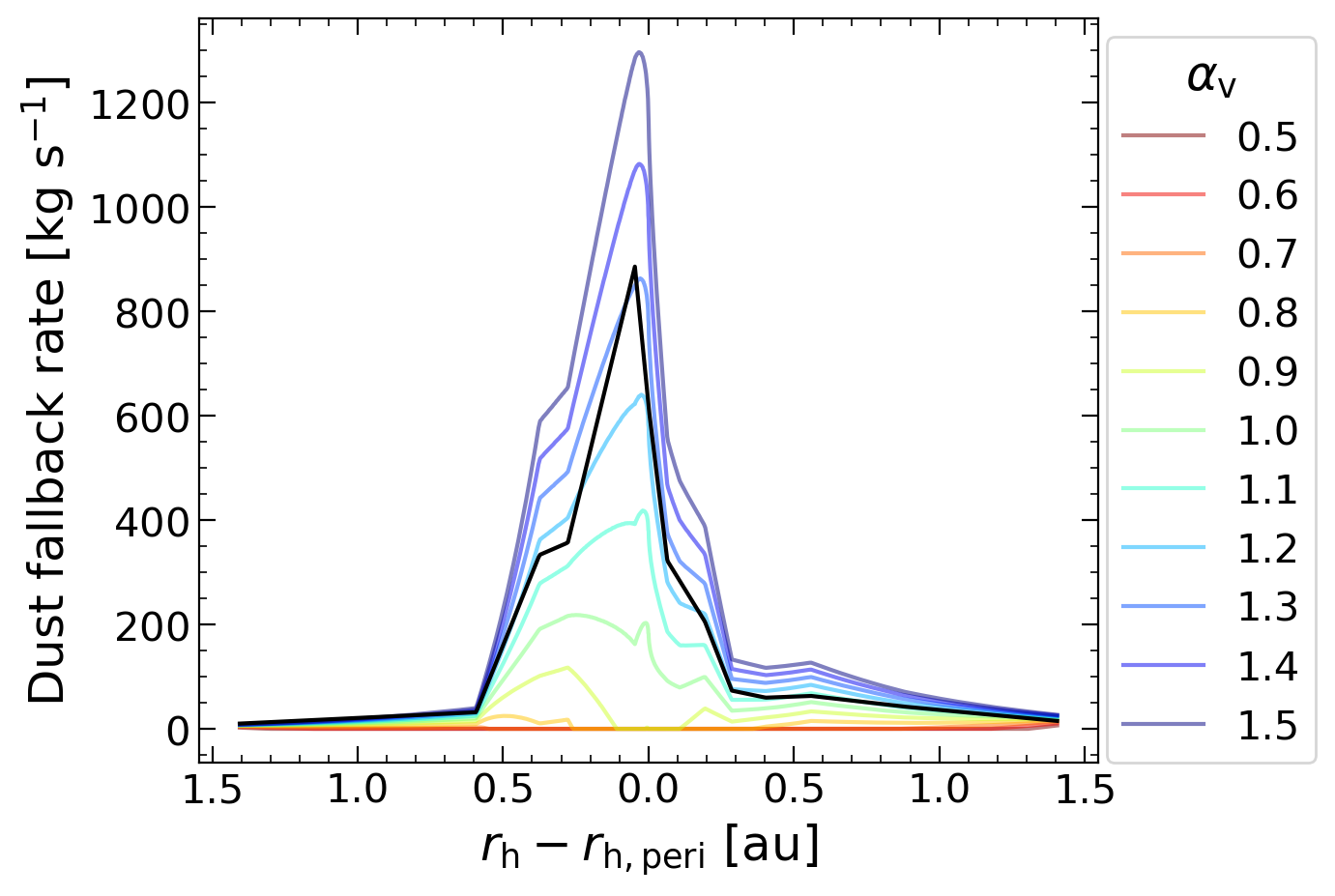}}
\caption{Dust fallback rate ($\dot{M}_{\rm{fallback}}$) with respect to the distance from the Sun and the $\alpha_{\rm{v}}$ values. The black solid line represents the dust mass escape rate adopted from \citet{Pozuelos2014A&A...571A..64P}.}
\label{fig_dust04}
\end{figure}

As expressed in Eq. (\ref{eq_dust06}), the amount of fallback debris inherently depends on the two edges of the fallback particle sizes, the maximum liftable size ($a_{\rm{max}}$) and the maximum escape size ($a_{\rm{escmax}}$), under the assumption of a single power law distribution. We believe that $a_{\rm{max}}$ can be determined solely by the balance of the local water sublimation environment and the nucleus surface gravity. We then calculated $a_{\rm{escmax}}$ by assuming the power index of the particle velocity distribution ($\alpha_{\rm{v}}$) at $R=20R_{\rm{N}}$ and using the observational constraints for the dust velocity along the normal direction ($v_{\rm{d}}$) of $a_0=1\ \rm{cm}$ particles at $R=20R_{\rm{N}}$ obtained by \citet{Pozuelos2014A&A...571A..64P}.

Fig. \ref{fig_dust03} shows the $a_{\rm{escmax}}$ values with respect to the distance from the Sun (with the perihelion distance subtracted, $r_\mathrm{h}$-$r_\mathrm{h, peri}$). As predicted from Fig. \ref{fig_dust02}, $a_{\rm{escmax}}$ increases as $\alpha_{\rm{v}}$ decreases. As an upper limit of $a_{\rm{escmax}}$, we plot the maximum liftable radius ($a_{\rm{max}}$) in Fig. \ref{fig_dust03}. $a_{\rm{escmax}}$ and $a_{\rm{max}}$ are derived by independent methods: $a_{\rm{escmax}}$ from the dust tail model in \citet{Pozuelos2014A&A...571A..64P} and the dust velocity distribution and $a_{\rm{max}}$ from the water sublimation model. In principle, the condition $a_{\rm{escmax}}=a_{\rm{max}}$ implies that all the particles ejected from the composite surface will escape the comet, so there is no fallback material (see the interval of the integral in Eq. (\ref{eq_dust06})).

Fig. \ref{fig_dust04} shows the dust fallback rate ($\dot{M}_{\rm{fallback}}$) with respect to the heliocentric distance and different $\alpha_{\rm{v}}$ values. For comparison, we plot the dust escape rate ($\dot{M}_{\rm{escape}}$) from \citet{Pozuelos2014A&A...571A..64P}. Due to the negative correlation between $a_{\rm{escmax}}$ and $\alpha_{\rm{v}}$, $\dot{M}_{\rm{fallback}}$ increases as $\alpha_{\rm{v}}$ increases. $\dot{M}_{\rm{fallback}}$ approaches zero for small $\alpha_{\rm{v}}$ ($\lesssim$ 1). Otherwise, for a large $\alpha_{\rm{v}}$ ($\gtrsim$ 1), $\dot{M}_{\rm{fallback}}$ with respect to the heliocentric distance is proportional to $\dot{M}_{\rm{escape}}$. The normalization coefficient ($k_{\rm{p}}$) in Eqs. (\ref{eq_dust05}) and (\ref{eq_dust06}) analytically explains this proportionality. A larger $\dot{M}_{\rm{escape}}$ leads to a larger $k_{\rm{p}}$, which simultaneously drives up $\dot{M}_{\rm{escape}}$ and $\dot{M}_{\rm{fallback}}$. More intuitively, an increase in dust production ($\dot{M}_{\rm{escape}}+\dot{M}_{\rm{fallback}}$) increases both $\dot{M}_{\rm{escape}}$ and $\dot{M}_{\rm{fallback}}$.

We integrated $\dot{M}_{\rm{fallback}}$ through one orbital revolution around the Sun. According to the result, $M_{\rm{fallback}}=4.06^{+14.84}_{-4.05}\times10^{9}\ \rm{kg}$, where we obtain the average at $\alpha_{\rm{v}}=1.0$, upper limit at $\alpha_{\rm{v}}=1.5$, and lower limit at $\alpha_{\rm{v}}=0.5$. Importantly, $\dot{M}_{\rm{fallback}}$ approaches zero for a small $\alpha_{\rm{v}}$. We compare $M_{\rm{fallback}}$ with the dust escape during one comet orbit ($M_{\rm{escape}}$) and obtain $M_{\rm{fallback}}/M_{\rm{escape}}=0.36^{+1.32}_{-0.36}$.

In principle, $v_{\rm{esc}}$ and $a_{\rm{max}}$ also depend on nucleus mass ($M_{\rm{N}}$), which has large uncertainty due to the poor constraint of the 81P bulk density \citep[$\rho_{\rm{bulk}}=200$--$800\ \rm{kg\ m^{-3}}$;][]{Davidsson2006Icar..180..224D, Szutowicz2008A&A...490..393S, Sosa2009MNRAS.393..192S}. Here we fixed $\rho_{\rm{bulk}}=600\ \rm{kg\ m^{-3}}$ and corresponding nucleus mass $M_{\rm{N}}=2.3\times10^{13}\ \rm{kg}$. The effect of nucleus bulk density on the fallback mass is discussed in Sect. \ref{subsec_EffectOfBulkDensity}.

\subsection{Excavation rate of 81P depressions}\label{subsec_ExcavationRate}

\begin{figure}
\resizebox{\hsize}{!}{\includegraphics{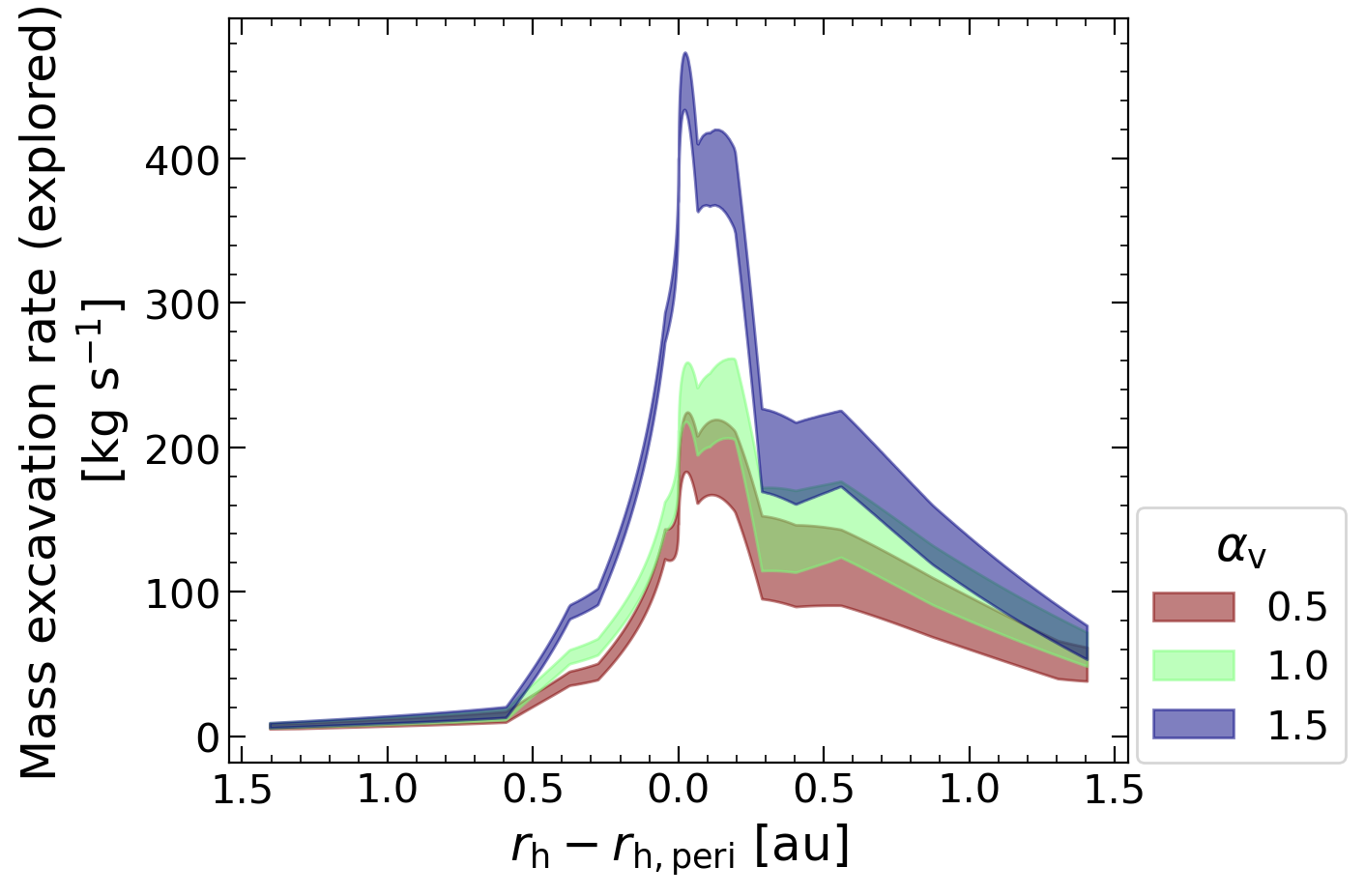}}
\caption{Mass excavation rate ($\dot{M}_{\rm{excavate}}$) from the explored side of 81P. Three cases of $\alpha_{\rm{v}}$ are presented. The filled range of each plot is attributed to the uncertainty in the water production rate.}
\label{fig_erosion01}
\end{figure}

\begin{table*}
\caption{\label{table_erosion01}Summary of our findings presented in Sects. \ref{sec_VolumeOfDepressions}--\ref{sec_DustEnv}.}
\centering
\setlength{\tabcolsep}{6pt}
\begin{tabular}{lccccccc}
\hline\hline
Region & $m_{\rm{H_2O}}Q_{\rm{H_2O}}$\tablefootmark{a} & $M_{\rm{escape}}$\tablefootmark{b} & $M_{\rm{fallback}}$\tablefootmark{c} & $M_{\rm{fallback}}/M_{\rm{escape}}$ & $M_{\rm{excavate}}$\tablefootmark{d} & $M_{\rm{depression}}$\tablefootmark{e} & $M_{\rm{excavate}} / M_{\rm{depreesion}}$ \\
& $[10^{9}\ \rm{kg}]$ & $[10^{9}\ \rm{kg}]$ & $[10^{9}\ \rm{kg}]$ & & $[10^{9}\ \rm{kg}]$ & $[10^{9}\ \rm{kg}]$ &  \\
\hline
entire   & $4.6$--$9.5$ & $11.3$ & $4.1^{+14.8}_{-4.1}$ & $0.36^{+1.32}_{-0.36}$ & $15.9$--$39.6$ & & \\ 
explored & $1.3$--$2.7$ & $2.3$  & $1.0^{+3.1}_{-1.0}$  & $0.36^{+1.32}_{-0.36}$ & $3.7$--$9.1$   & $174.5^{+15.0}_{-28.6}$ & $0.02$--$0.06$ \\
\hline
\end{tabular}
\tablefoot{
\tablefoottext{a}{Water production mass during one comet orbit.}
\tablefoottext{b}{Dust escape mass during one comet orbit \citep{Pozuelos2014A&A...571A..64P}.}
\tablefoottext{c}{Dust fallback mass during one comet orbit. The center values are decided when $\alpha_{\rm{v}}=1.0$, and the lower and upper limits are when $\alpha_{\rm{v}}=0.5$ and $\alpha_{\rm{v}}=1.5$, respectively.}
\tablefoottext{d}{Excavated mass during one comet orbit ($m_{\rm{H_2O}}Q_{\rm{H_2O}}+M_{\rm{escape}}+M_{\rm{fallback}}$).}
\tablefootmark{e}{Depressions mass converted from their volumes with $\rho_{\rm{bulk}}=600\ \rm{kg\ m^{-3}}$.}
}
\end{table*}

In this subsection, we summarize our results and compare the depression mass ($M_{\rm{depression}}$; Sect. \ref{sec_VolumeOfDepressions}) with the water sublimation ($Q_{\rm{H_2O}}$; Sect. \ref{sec_WaterEnv}) and dust ejection ($M_{\rm{escape}}+M_{\rm{fallback}}$; Sect. \ref{sec_DustEnv}) to estimate the "excavation rate" of the depressions. We express the mass excavation rate ($\rm{kg\ s^{-1}}$) of the nucleus surface as follows:
\begin{equation}\label{eq_erosion01}
    \dot{M}_{\rm{excavate}} = m_{\rm{H_2O}} Q_{\rm{H_2O}}+ \dot{M}_{\rm{escape}} + \dot{M}_{\rm{fallback}} ~.
\end{equation}
If we only consider the explored side of 81P, we can revise the above equation to
\begin{equation}\label{eq_erosion02}
    \dot{M}_{\rm{excavate}} = m_{\rm{H_2O}} [Q_{\rm{H_2O}}]_{\rm{explored}} + \left(\dot{M}_{\rm{escape}} + \dot{M}_{\rm{fallback}}\right)~ \frac{[Q_{\rm{H_2O}}]_{\rm{explored}}}{Q_{\rm{H_2O}}},
\end{equation}
where $[Q_{\rm{H_2O}}]_{\rm{explored}}$ is the water production rate on the explored side (the blue-filled region in Fig. \ref{fig_water01}). The second term of Eq. (\ref{eq_erosion02}) suggests that water gas lifts up dust in proportion to its sublimation activity.

Fig. \ref{fig_erosion01} shows the results for $\dot{M}_{\rm{excavate}}$ from the explored side (only three cases, namely, $\alpha_{\rm{v}}=0.5, 1.0, 1.5$, are presented in the plot). In case of $M_{\rm{N}}=2.3\times10^{13}\ \rm{kg}$ \citep[$\rho_{\rm{bulk}}=600\ \rm{kg\ m^{-3}}$;][]{Davidsson2006Icar..180..224D}, we constrain the total excavated mass from the explored side during one integrated orbit by $M_{\rm{excavate}} = 3.7$--$9.1 \times 10^9\ \rm{kg}$, where the lower limit occurs when $\alpha_{\rm{v}}=0.5$ and $f=0.08$ and the upper limit occurs when $\alpha_{\rm{v}}=1.5$ and $f=0.16$ (i.e., dependent on the dust fallback and water production). The estimated maximum rate of excavated mass is $\dot{M}_{\rm{excavate}}\sim250\ \rm{kg\ s^{-1}}$, occurring $\sim$20 days after the perihelion. Approximately 70 \% of the surface material is excavated after the perihelion due to asymmetric water production.

Comparing $M_{\rm{depression}}$ with $M_{\rm{excavation}}$ from the explored side enables us to determine the excavation rate of the 81P depressions. The result yields $M_{\rm{excavation}}/M_{\rm{depression}}=0.02$--$0.06$, implying 2--6 \% of the total depressions mass (or volume) on the explored side are excavated every comet orbit around the Sun. From the Jupiter encounter in 1974 to the Stardust encounter in 2004, 81P revolved five times, maintaining its constant JFC orbit and the activity level \citep{Sekanina2003JGRE..108.8112S}. Accordingly, we conclude that the total excavated mass of the 81P depressions due to the JFC-phase activity is up to 30 \% of the total depression mass.

We summarize our findings in Table \ref{table_erosion01}. Throughout the estimation, we fix the nucleus bulk density with $\rho_{\rm{bulk}}=600\ \rm{kg\ m^{-3}}$. The effect of bulk density on our results is discussed in Sect. \ref{subsec_EffectOfBulkDensity}. In addition, the estimate is an ideal upper limit of the excavation rate of the depressions, assuming all the water sublimation occurs inside the depressions. We further discuss the heterogeneity of the local excavation in Sect. \ref{subsec_LocationOfMassExcavation}, under the consideration of the 81P topography and the solar illumination conditions. 

\section{Discussion}\label{sec_discussion}

In this section, we discuss the validity, reliability, and novelty of the results by comparing them with previous research. We first discuss the novelty of this study in Sect. \ref{subsec_Novelty}. We discuss the consistency of 81P activity level during 30 years in Sect. \ref{subsec_ThirtyyearActivity}, the reliability and interpretability of our results on the number of depressions in Sect. \ref{subsec_DepressionsOn81P}, water environment in Sect. \ref{subsec_WaterEnvironment}, and fallback debris in Sect. \ref{subsec_FallbackDebris}. Then, we derive the dust-to-ice mass ratio range of 81P under a few assumptions in Sect. \ref{subsec_DTIRatio}. We discuss the effect of nucleus bulk density and the surface heterogeneity on our results in Sect. \ref{subsec_EffectOfBulkDensity} and Sect. \ref{subsec_LocationOfMassExcavation}, respectively. Finally, we discuss the implications of our results for the evolution and origin of 81P depressions in Sects. \ref{subsec_EnlargementOfDepressions}--\ref{subsec_OriginsOfDepressions}.

Throughout Sects. \ref{subsec_DepressionsOn81P}--\ref{subsec_DTIRatio}, we discuss our results on the depression count, water environment, fallback debris, and dust-to-ice mass ratio of 81P by comparing them with those for 67P. Both comets share similar orbital traits and histories, having comparable semimajor axes ($a=3.45\ \rm{au}$ for 81P and $a=3.46\ \rm{au}$ for 67P), perihelion distances ($q = 1.69\ \rm{au}$ for 81P and $q = 1.21\ \rm{au}$ for 67P), obliquity of the rotation pole axes ($I = 55\degree$ for 81P and $I=52\degree$ for 81P), and years of injection into the current JFC orbit (in 1974 for 81P and in 1959 for 67P). This comparative analysis enhances the understanding of the characteristics of 81P by leveraging insights gained from the well-studied comet 67P.

\subsection{Novelty of this study}\label{subsec_Novelty}

We conducted unique research by integrating observational evidence from 81P \citep{Pozuelos2014A&A...571A..64P} into the analysis of depression enlargement, addressing the potential limitations of previous research. In fact, \citet{GuilbertLepoutre2023PSJ.....4..220G} examined the depression enlargement of various JFCs. However, their approach relies on thermal parameters determined in 67P research; thus, it may overlook differences in the surface conditions of 81P and 67P. In contrast, our study incorporates observational constraints from \citet{Pozuelos2014A&A...571A..64P} when analyzing the enlargement of 81P depressions, representing the first attempt at treating comets other than 67P. Although our estimates of the dust environment of 81P (including escaping and fallback dust debris) may not be as accurate as in situ measurements of 67P \citep{Patzold2019MNRAS.483.2337P, Marschall2020FrP.....8..227M}, we provide reasonable results considering the broad range of parameters ($f$, $\alpha_{\rm{v}}$), as well as strong observational evidence from \citet{Pozuelos2014A&A...571A..64P}, who observed 81P across a wide range of orbital locations.

\subsection{Long-term activity of 81P}\label{subsec_ThirtyyearActivity}

Throughout Sects. \ref{sec_WaterEnv}--\ref{sec_DustEnv}, we used ground-based observational data on water and dust production rates to estimate the seasonal activity of 81P. For water production rates (Sect. \ref{sec_WaterEnv}), we relied on data from the 1997 and 2010 apparitions, when the visibility of the comet was favorable (Table \ref{table_water1}). Additionally, as described in Section \ref{subsec_DustEjectionModel}, our dust ejection model was developed using observational data from \citet{Pozuelos2014A&A...571A..64P}, who collected data across a wide range of heliocentric distances during the 2010 apparition.

We assumed that 81P's activity remained consistent across its orbits after it was injected into its current orbit in 1974. From 1974 to 2010, data on water and dust production are available only for the 1997 and 2010 apparitions. However, visual magnitude data for earlier apparitions were analyzed by \citet{Sekanina2003JGRE..108.8112S}. According to his study, the brightness of the comet was consistent during the first two apparitions (1978 and 1984) but dimmed by about one magnitude in the third apparition (1990). This decrease coincided with an increase in the perihelion distance, from $q=1.49\ \rm{au}$ to $q=1.58\ \rm{au}$, after a moderate encounter with Jupiter in 1987. However, \citet{Sekanina2003JGRE..108.8112S} noted that this change in brightness may not be directly related to the change in the perihelion distance, as the brightness returned to previous levels during the fourth apparition (1997), even though the perihelion distance remained unchanged. The comet was not observed during the fifth apparition (2003), but its activity was likely similar to previous levels, as the activity observed in the fourth (1997) and sixth (2010) apparitions was consistent \citep{deValBorro2010A&A...521L..50D}.

In summary, the overall activity level of 81P remained stable for the six apparitions from 1974 to 2010, despite a temporary brightness decrease in the 1990 apparition. Thus, we confidently extended our water and dust activity models based on the 1997 and 2010 data to earlier 81P orbits.

\subsection{Reliability of our depression count}\label{subsec_DepressionsOn81P}

We detected a total of 15 depressions. This number is less than the 23 depressions detected by \citet{Kirk2005LPI....36.2244K} in the same region. It is not clear why there is a discrepancy. We utilized the edge detection algorithm, while \citet{Kirk2005LPI....36.2244K} detected depressions by visual inspection of a false color elevation map, which made it difficult for us to trace their results. However, we ensured the inclusion of all the large depressions ($D\gtrsim500\ \rm{m}$) in our counting. \citet{Kirk2005LPI....36.2244K} estimated about 13 depressions with $D>500\ \rm{m}$, while we detected 13 depressions with $D\gtrsim400\ \rm{m}$ (Table \ref{table_result1}). On the other hand, small depressions with $D\lesssim400\ \rm{m}$ may have led to the discrepancy in counts between our study (2 depressions) and \citet{Kirk2005LPI....36.2244K} (about 10 depressions). Nevertheless, the volume of each depression with $D<400\ \rm{m}$ corresponds to $<$1 \% of our estimates ($<0.001\ \rm{km^3}$ / $0.290\ \rm{km^3}$). Therefore, we expect that uncounted small depressions contribute only to 10 \% of the total volume of depression. As seen in Sect. \ref{subsec_Result01}, the five largest depressions encompass the majority (90 \%) of the total volume of depression. Therefore, the discrepancy in the depression counting between this study and \citet{Kirk2005LPI....36.2244K} would not influence our main result, as it has minor contributions to the total volume of depression.

Although the total number of detected depressions differs between \citet{Kirk2005LPI....36.2244K} and this study, this study shows good agreement with \citet{Kirk2005LPI....36.2244K} regarding the depression diameters. They derived the dimensions of the depressions, with depths ranging from $50\ \rm{m}$ to $500\ \rm{m}$ and diameters ranging from $250\ \rm{m}$ to $2500\ \rm{m}$, and obtained a depth-to-diameter ratio of $d/D\sim0.2$. We derive $d/D\sim0.12$--$0.2$ from the depression diameter range of $100$--$1000$ m. We also find that the $d/D$ ratio in our research is consistent with those of 9P and 67P, most of which had ratios of $d/D\sim0.1$ \citep[9P,][]{Thomas2013Icar..222..453T} and $d/D=0.1$--$0.2$ \citep[67P,][]{Ip2016A&A...591A.132I}, respectively. Despite the small number of depression samples in our research, interestingly, the $d/D$ ratio is also similar to those of impact craters observed on other airless bodies, such as lunar craters \citep[$d/D=0.8$--$0.17$;][]{Wu2022GeoRL..4900886W} and Ryugu \citep[$d/D=0.09\pm0.02$;][]{Noguchi2021Icar..35414016N}. This similarity in the $d/D$ ratios may imply that we cannot rule out the impact origins of depressions on JFCs, although they have a distinct flat bottom and steep cliff morphology. The morphological trait suggests that the JFC depressions may have undergone significant modification from their original shapes \citep{Basilevsky2006P&SS...54..808B, Cheng2013Icar..222..808C}. Further discussion on the impact origin of these depressions is provided in Sect. \ref{subsubsec_ImpactCatering}.

\subsection{Asymmetric water production}\label{subsec_WaterEnvironment}

Our water sublimation model successfully reproduces the observed asymmetry in water production for the perihelion. For example, the combined observational data (depicted as black dots in Fig. \ref{fig_water02}) yield $Q_{\rm{H_2O}}=(9.3\pm2.0)\times10^{27}\ \rm{molecule\ s^{-1}}$ 25 days before perihelion and $Q_{\rm{H_2O}}=(5.6\pm0.5)\times10^{27}\ \rm{molecule\ s^{-1}}$ 25 days after perihelion. Our model closely matches these values to the accuracy of the observation, predicting $Q_{\rm{H_2O}}=8.4\times10^{27}\ \rm{molecule\ s^{-1}}$ and $Q_{\rm{H_2O}}=6.9\times10^{27}\ \rm{molecule\ s^{-1}}$ on the same days, assuming an average effective active fraction ($f=0.12$). While our model exhibits slightly smaller variations, it broadly aligns with the observational data, considering the associated error bars. The observed asymmetry has often been attributed to the seasonal effect of two confined jets positioned near the north pole (latitude $>75\degree$) and the southern hemisphere \citep{Farnham2005Icar..173..533F}. However, the exclusion of this effect in our model suggests that the asymmetry arises primarily from the shape of the nucleus, coupled with its high obliquity ($I=55\degree$). This result is consistent with earlier discussions regarding the asymmetric water production trend of JFCs (including 81P) \citep{Marshall2019A&A...623A.120M}.

The integrated water production over one cometary orbit indicates that the water production in the southern hemisphere is two to three times greater than that in the northern hemisphere. A critical aspect of this finding is that the explored side represents a relatively less eroded region. Similar disparities were reported in 67P with a high obliquity ($I = 52\degree$), where the southern side showed a factor of four greater erosion rate than did the northern side due to intensive solar insolation in the southern hemisphere near the perihelion \citep{Keller2015A&A...583A..34K, Thomas2015A&A...583A..17T}.

Our model estimates the effective active fraction within the range of $0.08<f<0.16$. This value is consistent with that of a previous study ($f=0.081$; \citealt{Jewitt2021AJ....161..261J}). By comparison, 67P exhibits $f\sim0.01$ under similar assumptions \citep{Marschall2016A&A...589A..90M} and $f\sim0.06$ under more realistic surface conditions \citep{Keller2015A&A...583A..34K}. Therefore, the surface of 81P appears more active than those of typical JFCs \citep{Sekanina2003JGRE..108.8112S, Pozuelos2014A&A...571A..64P}.

\subsection{Amount of fallback debris}\label{subsec_FallbackDebris}

\begin{table}

\caption{\label{table_fallback}Estimated fallback debris of 81P and 67P from our model.}
\centering
\renewcommand{\arraystretch}{1.1}
\begin{tabular}{cccc}
\hline\hline
$\alpha_{\rm{v}}$ & $M_{\rm{fallback,\ 81P}}$ & Thickness (81P) & $M_{\rm{fallback,\ 67P}}$ \\
 & [$10^9\ \rm{kg}$] & [cm] & [$10^9\ \rm{kg}$] \\
\hline
$0.5$ & $0.00$ & $0.0$ & $0.00$ \\
$0.6$ & $0.01$ & $0.0$ & $0.21$ \\
$0.7$ & $0.03$ & $0.1$ & $2.74$ \\
$0.8$ & $0.17$ & $0.5$ & $6.88$ \\ 
$0.9$ & $1.29$ & $3.8$ & $12.00$ \\ 
$1.0$ & $4.42$ & $13.1$ & $17.88$ \\
$1.1$ & $7.62$ & $22.6$ & $24.39$ \\
$1.2$ & $10.79$ & $32.0$ & $31.36$ \\
$1.3$ & $13.90$ & $41.3$ & $38.67$ \\
$1.4$ & $16.93$ & $50.3$ & $46.20$ \\
$1.5$ & $19.85$ & $59.0$ & $53.86$ \\
\hline

\end{tabular}
\tablefoot{$\rho_{\rm{bulk,\ 81P}}=600\ \rm{kg\ m^{-3}}$ is assumed.
}
\end{table}

Our investigation provides a constraint on the total fallback debris mass of $M_{\rm{fallback}}<2.0\times10^{10}\ \rm{kg}$ during one comet orbit (Table \ref{table_fallback}). Assuming a homogeneous distribution of the debris over the entire surface ($\sim56\ \rm{km^2}$), we infer that the debris would produce a dust layer of $\lesssim50\ \rm{cm}$ thickness per orbit around the Sun. Here, we posit that the bulk density of this dust layer equals the nucleus bulk density \citep[$\rho_{\rm{bulk}}=600\ \rm{kg\ m^{-3}}$;][]{Davidsson2006Icar..180..224D}. If we further assume that the debris accumulates solely on regions with low gravitational slope ($<30\degree$), the layer thickness is estimated to double ($\lesssim1\ \rm{m}$). Here, we postulate that the low gravitational slope region constitutes 50 \% of the entire surface based on the shape model of the explored side. The slope criterion considers the angle of repose of the debris \citep{Marschall2017A&A...605A.112M}.

This is the first attempt to constrain the amount of fallback debris other than for 67P \citep{Patzold2019MNRAS.483.2337P, Marschall2020FrP.....8..227M, Davidsson2021Icar..35414004D}. Despite the simplicity of our modeling compared to previous work, our estimate for the fallback debris aligns well with that of 67P. We constrain the ratio between fallback and escape to $M_{\rm{fallback}}/(M_{\rm{escape}}+m_{\rm{H_2O}} Q_{\rm{H_2O}})\sim1.2$ or less, which is similar to the 1.8--4.3 range found for 67P \citep{Patzold2019MNRAS.483.2337P}. The estimated thickness of the fallback debris layer also corresponds to the findings for 67P: \citet{Davidsson2021Icar..35414004D} estimated decimeter-scale dust deposits (0.1--1 m) per comet orbit depending on the location in the nucleus, and \citet{Marschall2020FrP.....8..227M} suggested a debris layer of $\lesssim$ 40 cm per comet orbit over the smooth regions on 67P.

However, the amount of fallback debris is strongly affected by $\alpha_{\rm{v}}$ and we cannot decide which value between $0.5<\alpha_{\rm{v}}<1.5$ is suitable for 81P. To retrieve the reliability of our dust model, we calculated the fallback debris of 67P adopting our model. The result is presented in the last column of Table \ref{table_fallback}. Overall, 67P shows approximately five times larger fallback debris compared to 81P, at the same $\alpha_{\rm{v}}$. Comparing our results with previous research, a sophisticated dust ejection model of \citet{Marschall2020FrP.....8..227M} estimated the amount of fallback debris of 67P during one comet orbit as $6.8^{+11}_{-6.8}\times10^8\ \rm{kg}$. Our result yields a similar amount of debris when $\alpha_{\rm{v}}$ falls between 0.6 and 0.7. If this value can be adaptable to 81P, the estimated fallback debris is less than $<0.1$ cm of deposits to the surface. However, due to the lack of observational constraints, we cannot decide which value is correct between this large uncertainty.

Our results suggest that 81P activity would add $\lesssim$ 5 m to the debris layer during the 30-year JFC phase. However, the actual distribution of the fallback debris may differ significantly from our simplified assumption. The spatial distribution of the debris may be heterogeneous, as observed on 67P, where much of the debris is transported from the southern hemisphere to the northern hemisphere during the polar night on the northern side near the perihelion \citep{Keller2015A&A...583A..34K, Thomas2015A&A...583A..17T, Lai2016MNRAS.462S.533L}. Local debris removal due to self-cleaning effects or intensive outbursts is also nonnegligible \citep{Davidsson2021Icar..35414004D}. Nevertheless, as pointed out by \citet{Davidsson2021Icar..35414004D}, similar mass transport (from the active south to the quiescent north) is unlikely to occur on 81P. Admittedly, our estimation is less sophisticated than those of well-studied 67P models. Nonetheless, we suggest that the actual fallback layer on 81P likely reached a few meters.

\subsection{Dust-to-ice mass ratio}\label{subsec_DTIRatio}

\begin{figure}
\resizebox{\hsize}{!}{\includegraphics{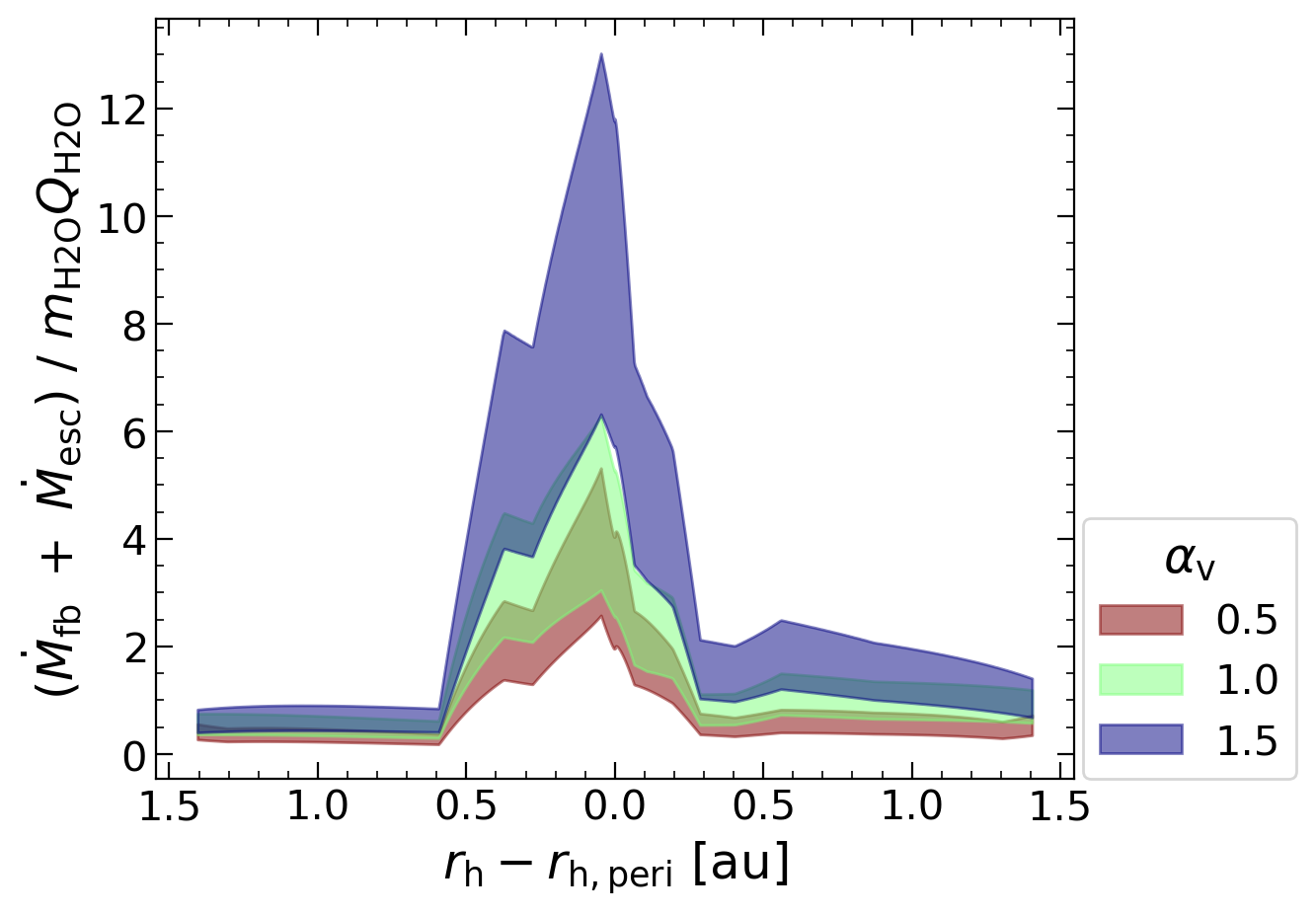}}
\caption{Dust-to-ice mass ratio with respect to the different heliocentric distances and $\alpha_{\rm{v}}$ values. The dust mass is the sum of the escaping and fallback masses.
}
\label{fig_DTIRatio}
\end{figure}

This research provides insights into a crucial parameter, namely, the mass ratio between dust and water ice, which is a fundamental characteristic of icy bodies. Determining this ratio is challenging. The conventional approach involves simultaneously measuring the production rates of escaping dust \citep[A$f\rho$;][]{AHearn1984AJ.....89..579A, Fink2012Icar..221..721F} and water ($Q_{\rm{H_2O}}$). However, this approach likely introduces bias, as the ratio varies with the heliocentric distance, with comae becoming more dusty at large distances \citep{AHearn1995Icar..118..223A}. Similarly, the average dust-to-ice mass ratio along the integrated orbit is likely biased.

As an alternative approach, including the dust fallback rate in the calculation would mitigate the potential bias. Our estimation yields a dust-to-ice mass ratio ($(M_{\rm{fallback}} + M_{\rm{escape}}) /m_{\rm{H_2O}} Q_{\rm{H_2O}}$) between 1 and 7. A similar approach suggests a similar value ($<$ 2.5) for 67P \citep{Marschall2020FrP.....8..227M}. However, as shown in Fig. \ref{fig_DTIRatio}, this ratio depends on the orbital location (especially the heliocentric distance), yielding values smaller than unity at large heliocentric distances and values substantially larger than unity near the peak production level (i.e., near the perihelion).

We attribute the heliocentric distance dependency of the dust-to-ice mass ratio in Fig. \ref{fig_DTIRatio} to two primary factors. First, dust ejection from the surface is contingent upon the gas sublimation environment, which determines the total dust amount ($M_{\rm{fallback}} + M_{\rm{escape}}$). The maximum liftable size ($a_{\rm{max}}$) is a function of the local gas sublimation rate ($Q_{\rm{H_2O}}$; Eq. (\ref{eq_dust02})). At a large heliocentric distance, $a_{\rm{max}}$ would be small, resulting in most of the remaining dust not being lifted. Second, the fallback "dust" ($\dot{M}_{\rm{fallback}}$) consists of not only refractory but also volatile ice. Evidence of this was observed during the Rosetta mission, where smooth terrains consisting of fallback debris frequently showed ice sublimation activity \citep{Hu2017A&A...604A.114H}. Furthermore, \citet{Davidsson2021Icar..35414004D} conducted a thermophysical simulation of ejected dust and concluded that decimeter-sized particles (the typical upper limit for fallback debris) retain $>$ 95 \% of water ice during the first 12 hours post-ejection, with even cm-sized particles preserving $>$ 40 \% of initial water ice. This could potentially introduce a bias in Fig. \ref{fig_DTIRatio}, overestimating the refractory dust mass, as sublimation from these ejected particles would still contribute to the total water production rate.

Moreover, our model does not rigorously account for the long-term evolution of the dust-to-ice mass ratio. Fallback dust deposits are known to suppress activity over time \citep{Attree2018A&A...610A..76A, Attree2023A&A...670A.170A}, leading to a gradual reduction in activity levels and an increase in the dust-to-ice mass ratio near the surface. This retroactive effect, which is absent from our model, could introduce further bias. The magnitude of this bias depends on the amount of fallback debris generated per comet orbit (see Sect. \ref{subsec_FallbackDebris}). If the dust layer deposits are less than $<1$ cm per orbit ($\alpha_{\rm{v}}\lesssim1$), the dust-to-ice mass ratio will evolve slowly, and the current values will remain close to the previous estimates. However, if the dust layer deposits exceed $>10$ cm per orbit ($\alpha_{\rm{v}}\gtrsim1$), the ratio will increase rapidly, introducing substantial bias into current estimates.

In short, it would be difficult to derive the true surface dust-to-ice mass ratio without comprehensive in situ measurements, as were obtained by the Rosetta mission. Nevertheless, we suggest that the dust-to-ice mass ratio of 81P near the perihelion (i.e., the peak value in Fig. \ref{fig_DTIRatio}) could reflect a close approximation to the true ratio. This approach mitigates the biases caused by the factors mentioned above. First, $a_{\rm{max}}$ increases near the perihelion due to intensive gas sublimation, incorporating most of the surface dust components into the mass calculation. Our estimation indicates a maximum liftable dust size of $a_{\rm{max}}\sim50\ \rm{cm}$ near the perihelion. Second, the volatile loss of the lifted dust would be maximized because of the intensive solar radiation and the longer flight time within the coma due to the high ejection velocity. As a result, we suggest that the surface dust-to-ice mass ratio of 81P falls between 2 and 14 depending on the water production efficiency ($f$) and the velocity distribution of the ejected dust ($\alpha_{\rm{v}}$). This result implies a dusty surface environment of 81P, which is consistent with that of 67P \citep[3--7.5;][]{Choukroun2020SSRv..216...44C}.

\subsection{Effect of bulk density}\label{subsec_EffectOfBulkDensity}

\begin{table*}
\caption{\label{table_erosion2}Effect of bulk density in our findings.}
\centering
\begin{tabular}{cccccccc}

\hline\hline
$\rho_{\rm{bulk}}$ & \it{M}$_{\rm{N}}$\tablefootmark{a} & \it{M}$_{\rm{depression}}$\tablefootmark{b} & \it{M}$_{\rm{fallback}}$\tablefootmark{c} & \it{M}$_{\rm{excavate}}$\tablefootmark{d} & \it{M}$_{\rm{fallback}}$/\it{M}$_{\rm{escape}}$ & \it{M}$_{\rm{excavate}}$/\it{M}$_{\rm{depression}}$ & DTI ratio\tablefootmark{e} \\

[$\rm{kg}$ $\rm{m^{-3}}$] & [$10^{13}$ kg] & [$10^9$ kg] & [$10^9$ kg] & [$10^9$ kg] & & & \\
\hline
$200$ & $0.76$ & $58.16^{+4.98}_{-9.53}$ & $11.14^{+27.83}_{-11.14}$ & $3.68$--$13.63$ & $<3.46$ & $0.06$--$0.28$ & $2.6$--$21.0$ \\
$400$ & $1.52$ & $116.32^{+9.97}_{-19.05}$ & $6.36^{+18.78}_{-6.35}$ & $3.68$--$10.53$ & $<2.23$ & $0.03$--$0.11$ & $2.6$--$15.5$ \\
$600$ & $2.28$ & $174.48^{+14.95}_{-28.58}$ & $4.06^{+14.84}_{-4.05}$ & $3.68$--$9.14$ & $<1.68$ & $0.02$--$0.06$ & $2.6$--$13.0$ \\
$800$ & $3.04$ & $232.64^{+19.93}_{-38.11}$ & $2.61^{+12.53}_{-2.60}$ & $3.68$--$8.31$ & $<1.34$ & $0.01$--$0.04$ & $2.6$--$11.5$ \\ 
\hline
\end{tabular}
\tablefoot{
\tablefoottext{a}{Nucleus mass with a given bulk density and ellipsoid-like shape \citep[2.75 $\times$ 2.00 $\times$ 1.65 km;][]{Duxbury2004JGRE..10912S02D}.}
\tablefoottext{b}{Total depression mass converted from their volumes with a given bulk density.}
\tablefoottext{c}{Dust fallback mass during one comet orbit. The central values are determined when $\alpha_{\rm{v}}=1.0$, and the lower and upper values are obtained assuming $\alpha_{\rm{v}}=0.5$ and $\alpha_{\rm{v}}=1.5$, respectively.} 
\tablefoottext{d}{Total excavated mass during one comet orbit from the explored side (Eq. (\ref{eq_erosion02})).}
\tablefoottext{e}{Dust-to-ice mass ratio.}
}
\end{table*}

Throughout this study, we have assumed a bulk density for 81P of $\rho_{\rm{bulk}}=600\ \rm{kg\ m^{-3}}$. However, estimates of the bulk density of 81P widely vary between different studies. \citet{Davidsson2006Icar..180..224D} proposed a bulk density of $\rho_{\rm{bulk}}<600$--$800\ \rm{kg\ m^{-3}}$, while \citet{Szutowicz2008A&A...490..393S} suggested $\rho_{\rm{bulk}}=400\pm200\ \rm{kg\ m^{-3}}$, and \citet{Sosa2009MNRAS.393..192S} estimated $\rho_{\rm{bulk}}=300^{+500}_{-300}\ \rm{kg\ m^{-3}}$. As a result, this introduces a large uncertainty not only in the estimation of depression mass ($M_{\rm{depression}}$) but also in the total nucleus mass. This, in turn, affects key parameters in our dust ejection model, particularly in calculations of the maximum escape radius ($a_{\rm{escmax}}$), the maximum liftable particle size ($a_{\rm{max}}$), and the amount of fallback debris ($M_{\rm{fallback}}$).

The effect of the bulk density on our findings is summarized in Table \ref{table_erosion2}. The estimated bulk density of 81P from previous studies lies between $\rho_{\rm{bulk}}=200$--$800\ \rm{kg\ m^{-3}}$, therefore, we considered four different values, $\rho_{\rm{bulk}}=200,\ 400,\ 600,$ and $800\ \rm{kg\ m^{-3}}$, and calculated the corresponding $M_{\rm{depression}}$ and $M_{\rm{fallback}}$. It is clear that $M_{\rm{depression}}$ increases linearly with the bulk density. In contrast, $M_{\rm{fallback}}$ decreases as the bulk density increases due to the reduction in $a_{\rm{max}}$, which limits the contribution of large dust particles to the total dust mass. The central value of $M_{\rm{fallback}}$ is determined when $\alpha_{\rm{v}}=1.0$, with the lower and upper limits defined by $\alpha_{\rm{v}}=0.5$ and $1.5$, respectively. As noted in Sect. \ref{subsec_Result03}, no fallback material is present in our dust model when $\alpha_{\rm{v}}=0.5$. In addition, in all cases, $M_{\rm{fallback}}$ is highly sensitive to the choice of $\alpha_{\rm{v}}$.

Variations in bulk density affect both $M_{\rm{depression}}$ and excavated mass ($M_{\rm{excavate}}$) from the explored side. Consequently, a wide range of "depression excavation rates" ($M_{\rm{excavate}}/M_{\rm{depression}}$) is possible, ranging from 1 \% to 28 \% of the depressions excavated per comet orbit. If we exclude the cases of very low bulk density ($\rho_{\rm{bulk}}=200\ \rm{kg\ m^{-3}}$), "typical" $M_{\rm{excavate}}/M_{\rm{depression}}$ falls between 1 \% and 11 \% per comet orbit. Although we cannot completely rule out the case of low bulk density, we conclude that the total volume excavated from the 81P depressions during the JFC phase is less than $\lesssim50$ \% of the depression volume (considering that 81P completed five revolutions around the Sun for 30 years).

This estimate could be an ideal (i.e., oversimplified) upper limit of the depression excavation rates, assuming that all water sublimation occurs inside the depressions. We further discuss the excavated volume of the depression in Sect. \ref{subsec_LocationOfMassExcavation}, considering the heterogeneous activity of the nucleus.

\subsection{Effect of activity location on the nucleus}\label{subsec_LocationOfMassExcavation}

\begin{figure*}
\centering

\includegraphics[width=17cm]{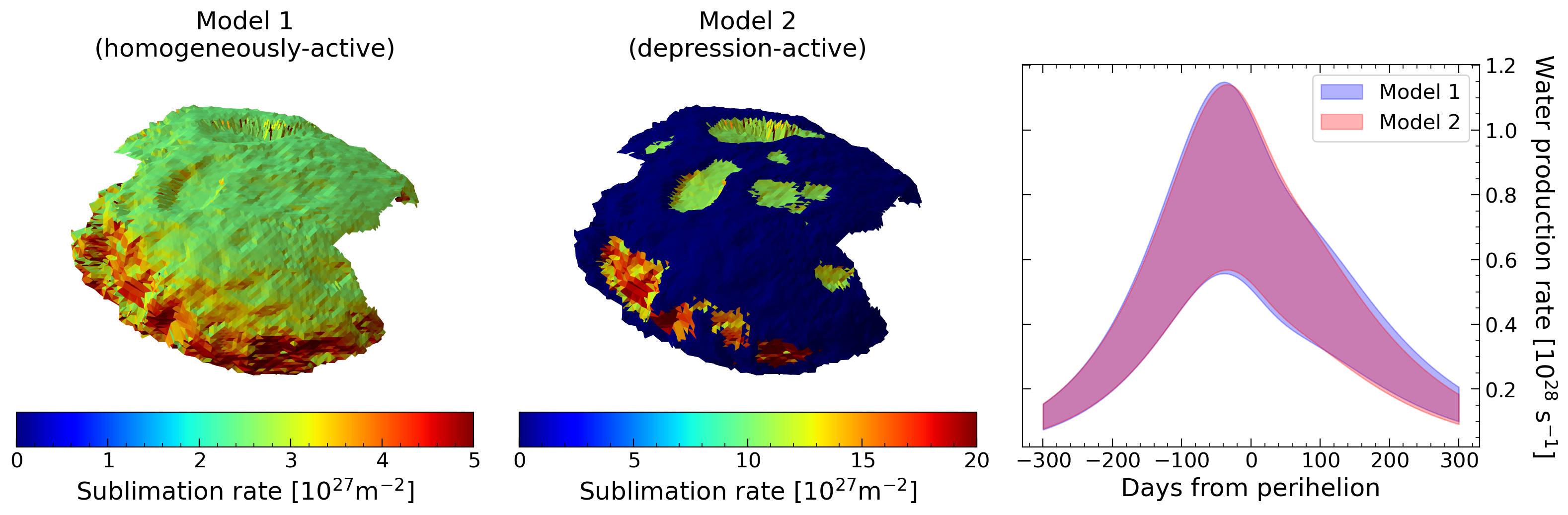}
\caption{Water sublimation model from different activity models. (left \& center) Orbit-integrated water sublimation of 81P’s explored side. The left shows a fully homogeneous model, and the center shows a depression-active model, respectively. (right) Time-dependent water production rates of two models.}

\label{fig_diffmodel}
\end{figure*}

In Sect. \ref{subsec_ExcavationRate}, we compared the depression volume with the surface excavated volume to estimate the depression excavation rate during the 30-year JFC phase. Our results constrain an upper limit for the depression excavation rate, with $M_{\rm{excavate}}/M_{\rm{depression}} <$ 6 \% per orbit and $<$ 30 \% over the 30-year JFC phase. This assumes that all excavations occurred \textit{inside} the depressions. Admittedly, our study does not rigorously define the exact location of the excavation on the nucleus surface, suggesting that not all sublimation-driven activity necessarily contributes to depression enlargement. Therefore, our estimate represents an upper limit for depression enlargement.

We assumed homogeneous activity across the surface, modeling water sublimation with a single scaling factor $f$ (Eq. (\ref{eq_water04})). However, this assumption overlooks the potential for a heterogeneous activity across the nucleus, where different surface regions may have different $f$ values, as found on the 67P's surface \citep[i.e., $f$ value has regional dependency;][]{Marschall2020FrP.....8..227M, Attree2023A&A...670A.170A}. Identifying these variations is challenging for 81P. In Fig. \ref{fig_diffmodel}, we present two extreme activity models: one assuming homogeneously active regions (our default assumption) and the other assuming only facets within the depressions are active (the "depression-active" model). In the latter case, the faces outside the depressions are assumed to have $f=0$, and the $f$ values inside the depressions are determined to fit the observational data. However, no significant differences are observed between the models. This highlights a limitation of our research because the available ground-based observational data are insufficient to precisely pinpoint the locations of activity on the nucleus.

Two primary factors contribute to this limitation. First, the depressions on 81P are evenly distributed on the explored side, resulting in minimal variation in overall activity trends. Second, the explored side of 81P contributes only a small fraction of total water production, while the unexplored southern hemisphere plays a more substantial role, especially near perihelion.

We further explored whether the degree of erosion could be different inside and outside the depressions. Two main factors likely drive heterogeneous erosion: solar energy input and gravitational slope. Solar energy input is the primary driver of heterogeneous activity \citep{Kossacki2019Icar..319..381K, Benseguane2022A&A...668A.132B} and is influenced by the oblique rotation pole \citep{Marshall2019A&A...623A.120M} and local topography \citep{Kossacki2019Icar..319..381K, Benseguane2022A&A...668A.132B}. For 81P, these effects are pronounced due to its large obliquity \citep[$I=55\degree$;][]{Sekanina2004Sci...304.1769S} and rough topography \citep{Brownlee2004Sci...304.1764B, Vincent2017MNRAS.469S.329V}. Gravitational slope also plays a role in determining the thickness of local refractory dust deposits, which can hinder solar heat transfer and reduce local sublimation activity. As shown by \citet{Hu2017A&A...604A.114H}, even a 10 mm thick dust layer can reduce activity to 20 \% of that on clean surfaces (without dust deposits). Our calculations suggest that fallback debris could accumulate to thicknesses exceeding 1 m in low-sloped regions (see Sect. \ref{subsec_FallbackDebris}), potentially suppressing subsurface sublimation.

However, according to our calculation, there are little differences inside and outside the depressions, in terms of the total solar energy input and the fraction of high gravitational slope ($>30\degree$). The average solar energy input is similar for the facets inside ($3.4\ \rm{GJ\ m^{-2}\ {orbit^{-1}}}$) and outside ($3.3\ \rm{GJ\ m^{-2}\ {orbit^{-1}}}$) depressions. Moreover, the area fraction with a gravitational slope larger than $>30\degree$ is the same for inside and outside (55 \%). In conclusion, the regions inside and outside the depressions do not exhibit significant differences in surface excavation, indicating nearly homogeneous erosion across the surface. Based on their surface area fraction, depressions are estimated to contribute approximately 40 \% of the total excavation. Accordingly, the estimated depression excavation rate is $M_{\rm{total}}/M_{\rm{dep}} < 3 \%$ per orbit and less than 15 \% over the 30-year JFC phase.

\subsection{Implications and potential caveats of depression enlargement}\label{subsec_EnlargementOfDepressions}

Our estimate of the excavated volume (in comparison with the depression volume) implies that the depressions on 81P barely expanded in diameter over its 30 years as a JFC. We calculate the excavation volume over 30 years, compare it with the depression diameters and volumes, and find that the diameters have increased by only approximately 1 \% of the original diameters (2--20 m). Our estimate is roughly consistent with the maximum erosion of the depressions of 81P determined in a previous study \citep[28 m;][]{GuilbertLepoutre2023PSJ.....4..220G}. This consistency supports the results of this previous numerical study, although their approach (a numerical approach based on thermal evolution) was different from ours (mostly based on the observational evidence of dust and gas ejections).

However, we recognize that our constraint does not fully account for the potential enlargement of the depressions. For example, cliff collapse can enlarge the depression diameter without a significant volume loss by accumulating collapsed chunks on depression bottoms. Evidence of cliff collapse and a subsequent increase in the number of boulders below the cliff was observed for 67P \citep{Groussin2015A&A...583A..32G, Pajola2015A&A...583A..37P, Davidsson2024MNRAS.527..112D}. Some observational evidence on 81P, such as the talus found in Left Foot, suggests that cliff collapses might have occurred in past orbits \citep{Brownlee2004Sci...304.1764B}. However, while such accumulation processes can increase the depression diameter, they do not result in volume excavation. Therefore, our analysis of depression "enlargement" pertains to depression volume but may not fully encompass changes in depression size due to these unknown enlargement mechanisms.

\subsection{Implications for the origins of 81P depressions}\label{subsec_OriginsOfDepressions}

Because sublimation-driven activity contributes less to depression enlargement according to our research, there should be other factors involved in this phenomenon. Here, we discuss two potential mechanisms, namely, cometary outbursts (Sect. \ref{subsubsec_Outburst}) and impact cratering (Sect. \ref{subsubsec_ImpactCatering}), as alternative origins of depressions on 81P and other JFCs.

\subsubsection{Outbursts}\label{subsubsec_Outburst}

Outbursts are observed across various comets, displaying a wide range of magnitudes and frequencies. However, their contribution to the formation of depressions on 81P remains uncertain. \citet{Pozuelos2014A&A...571A..64P} observed two outbursts during the 2010 apparition, estimating the total mass ejected from these events as $M_{\rm{ej}}\sim9.2 \times 10^8$ kg $M_{\rm{ej}}\sim3.0 \times 10^8$ kg. Assuming these outbursts excavated surface mass (with $\rho_{\rm{bulk}}=600\ \rm{kg\ m^{-3}}$; \citealt{Davidsson2006Icar..180..224D}) and that all the excavated mass escaped from the nucleus, the corresponding excavated volume is approximately $V\sim0.001\ \rm{km^3}$ for each outburst. The total excavated volume from these two outbursts is thus less than 1 \% of the total depression volume on the explored side ($0.288\ \rm{km^3}$). Obviously, the outbursts observed by \citet{Pozuelos2014A&A...571A..64P} in the 2010 apparition are not associated with the depressions observed during the Stardust encounter in 2004. Therefore, it is unlikely that the outbursts strongly contributed to the formation of depressions within the 30-year JFC orbit. Although we definitively exclude the possibility that 81P underwent large-scale outbursts after it became a JFC to create depressions on the surface, the possibility of outbursts at large heliocentric distances cannot be ruled out \citep{Gronkowski2016EM&P..119...23G}.

\subsubsection{Impact cratering}\label{subsubsec_ImpactCatering}

\begin{figure}
\centering
\resizebox{\hsize}{!}{\includegraphics{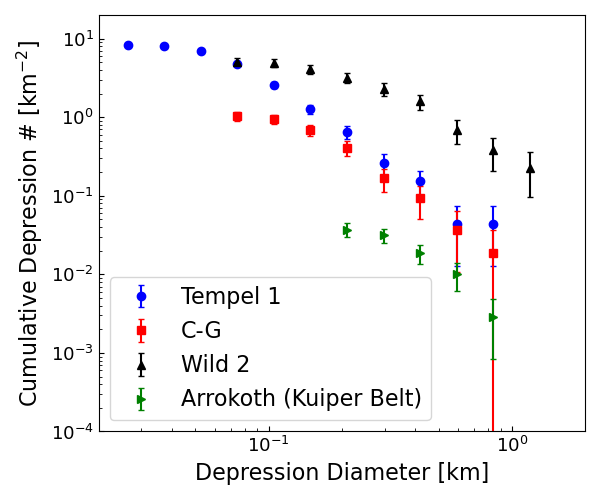}}
\caption{Cumulative depression size distributions of three JFCs (9P, 67P, and 81P) and one Kuiper Belt Object Arrokoth. The data used were obtained from the following studies: 9P---\citet{Thomas2013Icar..222..453T}; 67P---\citet{Ip2016A&A...591A.132I}; 81P---\citet{Kirk2005LPI....36.2244K}; and Arrokoth---\citet{Spencer2020Sci...367.3999S}.}
\label{fig_CSD}
\end{figure}

Impact cratering is one of the most significant surface modification processes for Solar System bodies, including comets. Given the short dynamic lifetime of JFCs ($<10^4$ years; \citealt{Levison1997Icar..127...13L}), it is unlikely that impact cratering formed dense depressions during the JFC orbit \citep{Vincent2015P&SS..107...53V}. Similarly, the contribution of impact cratering during the Centaurs (i.e., pre-JFC) phase is negligible compared to the Kuiper Belt, given their short dynamic lifetime and low impact probability \citep{Durda2000Icar..145..220D}.

\citet{Ip2016A&A...591A.132I} investigated depressions on three JFCs (9P, 67P, and 81P) and find that they share a similar power-law slope in the depression size distribution, implying that the depressions commonly formed before they became JFCs. These slopes are also consistent with the impact craters on Arrokoth, the Kuiper Belt Object (Fig. \ref{fig_CSD}). This may imply that the depressions on the three JFCs formed during the long-term residence in their reservoir. However, direct comparison of size distributions between JFCs and Arrokoth should be treated with caution, since JFCs likely originated from different subpopulations with Arrokoth \citep{Nesvorny2017ApJ...845...27N}. Furthermore, the surface densities of the JFC depressions are one to two orders of magnitude higher than those of the Arrokoth depressions. To account for this excess, 10 to 100 times more frequent impacts must have occurred compared to the expected impact rates on Kuiper Belt Objects \citep{DellOro2013A&A...558A..95D, Abedin2021AJ....161..195A}. It is not certain whether such a diversity of crater densities occurs in the Kuiper Belt region while forming a similar size distribution for the different subpopulations, which requires further investigation.

\section{Conclusions and summary}\label{sec_conclusion}

We studied the water and dust environment of 81P (water escape, dust escape, and dust fallback) and estimated the excavation rates of depressions, comparing them with the depression volume. We conjecture that the depressions were preserved during the 30-year JFC phase. Below, we summarize our main results.

\begin{enumerate}

    \item The total excavated mass (volume) on the explored side of 81P is $3.7$--$9.1\times10^9\ \rm{kg}$ ($0.006$--$0.015\ \rm{km^3}$) during one comet orbit around the Sun. This constrains the maximum excavation rate of the depression ($0.288\ \rm{km^3}$) to within 5 \% per one comet orbit and less than 30 \% during the 30-year JFC phase.

    \item We constrain the total amount of fallback debris to $M_{\rm{fallback}}<2.0\times10^{10}\ \rm{kg}$ during one comet orbit around the Sun. The expected thickness of the debris is $\lesssim50\ \rm{cm}$ in one comet orbit, assuming a homogeneous distribution of the debris. This result is consistent with that for 67P.

    \item We constrain the dust-to-ice mass ratio of the 81P surface to between 2 and 14 depending on the water production efficiency ($f$) and the velocity distribution of the ejected dust ($\alpha_{\rm{v}}$). This result is consistent with that for 67P.

\end{enumerate}

\begin{acknowledgements}
This research was supported by a National Research Foundation of Korea (NRF) grant funded by the Korean government (MEST) (No. 2023R1A2C1006180). We also appreciate the anonymous reviewer for their careful reading and insightful comments.
\end{acknowledgements}

\bibliographystyle{aa} 

\begin{appendix}

\onecolumn
\renewcommand{\arraystretch}{1.1}

\section{Additional materials}\label{appendix_material}
\begin{table}[!htbp]
\caption{\label{table_water1}Water production rates of 81P observed in the 1997 and 2010 apparitions.}
\centering

\begin{tabular}{cccccccc}
\hline\hline
Days from perihelion & Apparition & $r_{\rm{h}}$ & $Q_{\rm{H_2O}}$ & $\sigma_{\rm{H_2O}}$ & $Q_{50}$\tablefootmark{a} & $\sigma_{50}$\tablefootmark{b} & Reference \\

[days] & [year] & [au] & [$10^{28}\ \rm{s^{-1}}$] & [$10^{28}\ \rm{s^{-1}}$] & [$10^{28}\ \rm{s^{-1}}$] & [$10^{28}\ \rm{s^{-1}}$] & \\
\hline
$-173$ & 1997 & 2.25 & $0.43$ & $0.02$ & $0.43$ & $0.00$ & 1 \\
\hline
$-110$ & 1997 & 1.91 & $0.40$ & $0.04$ & $0.42$ & $0.05$ & 1 \\
$-110$ & 1997 & 1.91 & $0.48$ & $0.06$ & & & 1 \\
$-108$ & 1997 & 1.90 & $0.44$ & $0.32$ & & & 1 \\
\hline
$-94$ & 1997 & 1.83 & $0.99$ & $0.24$ & $0.73$ & $0.09$ & 2 \\
$-93$ & 1997 & 1.82 & $0.69$ & $0.02$ & & & 1 \\
$-84$ & 1997 & 1.78 & $0.72$ & $0.02$ & & & 1 \\
$-84$ & 1997 & 1.78 & $0.91$ & $0.04$ & & & 1 \\
$-81$ & 1997 & 1.77 & $1.00$ & $0.07$ & & & 1 \\
$-81$ & 1997 & 1.77 & $1.15$ & $0.26$ & & & 1 \\
$-66$ & 1997 & 1.71 & $1.58$ & $0.21$ & & & 2 \\
$-62$ & 1997 & 1.70 & $0.69$ & $0.33$ & & & 1 \\
\hline
$-35$ & 1997 & 1.62 & $1.51$ & $0.32$ & $0.93$ & $0.20$ & 2 \\
$-21$ & 2010 & 1.61 & $1.00$ & $0.05$ & & & 3 \\
$-20$ & 2010 & 1.61 & $1.07$ & $0.06$ & & & 3 \\
$-20$ & 2010 & 1.61 & $0.86$ & $0.05$ & & & 3 \\
$-19$ & 2010 & 1.61 & $1.21$ & $0.15$ & & & 4 \\
$-18$ & 2010 & 1.61 & $1.13$ & $0.06$ & & & 3 \\
$-9$ & 1997 & 1.58 & $1.38$ & $0.43$ & & & 2 \\
$-8$ & 1997 & 1.58 & $0.65$ & $0.08$ & & & 1 \\
$-8$ & 1997 & 1.58 & $0.66$ & $0.12$ & & & 1 \\
$-8$ & 1997 & 1.58 & $0.47$ & $0.11$ & & & 1 \\
$-1$ & 2010 & 1.60 & $0.99$ & $0.16$ & & & 5 \\
$-1$ & 2010 & 1.60 & $0.75$ & $0.12$ & & & 5 \\
\hline
$+23$ & 1997 & 1.60 & $0.85$ & $0.34$ & $0.56$ & $0.05$ & 2 \\
$+28$ & 1997 & 1.61 & $0.51$ & $0.04$ & & & 1 \\
$+28$ & 1997 & 1.61 & $0.54$ & $0.04$ & & & 1 \\
$+30$ & 1997 & 1.61 & $0.60$ & $0.03$ & & & 1 \\
$+33$ & 2010 & 1.63 & $0.52$ & $0.08$ & & & 5 \\
$+34$ & 2010 & 1.63 & $0.52$ & $0.08$ & & & 5 \\
\hline
$+56$ & 1997 & 1.68 & $0.45$ & $0.05$ & $0.46$ & $0.09$ & 1 \\
$+56$ & 1997 & 1.68 & $0.58$ & $0.06$ & & & 1 \\
$+56$ & 1997 & 1.68 & $0.40$ & $0.05$ & & & 1 \\
$+58$ & 1997 & 1.68 & $0.50$ & $0.04$ & & & 1 \\
$+58$ & 1997 & 1.68 & $0.32$ & $0.06$ & & & 1 \\
\hline
\end{tabular}

\tablefoot{
\tablefoottext{a}{Eq. (\ref{eq_water05})}
\tablefoottext{b}{Eq. (\ref{eq_water06})}
}

\tablebib{
(1) \citet{Farnham2005Icar..173..533F}; (2) \citet{Fink1999Icar..141..331F}; (3) \citet{deValBorro2010A&A...521L..50D}; (4) \citet{Hashimoto2012PASJ...64...27H}; (5) \citet{DelloRusso2014Icar..238..125D}.
}

\end{table}

\end{appendix}

\end{document}